%% file: letter.tex
\documentclass[11pt,a4paper]{article}

\usepackage{jheppub}

\usepackage{rotating}
\usepackage{epstopdf}
\usepackage{overpic}
\usepackage{subfigure}
\usepackage{array}
\usepackage[section]{placeins}
\usepackage{chngpage}
\usepackage{hyperref} 
\usepackage{multirow}
\usepackage{setspace}
\usepackage{atlasphysics} 
\usepackage{lmetdefs} 
\usepackage{microtype}  
\usepackage{bm}

\newcommand{\QCDBg}{multi-jet background}

\usepackage{preprintcover}  

\PreprintCoverPaperTitle{
Search for new particles in events with one lepton and missing transverse momentum
in $pp$ collisions at $\sqrt{s}$~=~8\, \tev\ with the ATLAS detector
} 

\PreprintIdNumber{
CERN-PH-EP-2014-139
} 

\PreprintCoverAbstract{
This paper presents a search for new particles in events with one lepton (electron or muon) and 
missing transverse momentum using 20.3~\ifb\  of proton--proton collision data at  
$\sqrt{s}=8\tev$ recorded by the ATLAS experiment at the Large Hadron Collider.
No significant excess beyond Standard Model expectations is observed.
A \wp\ with Sequential Standard Model couplings is excluded at the 95\% confidence level for masses up to 3.24~\tev.
Excited chiral bosons (\wstar) with equivalent coupling strengths are excluded for masses up to 3.21~\tev.
In the framework of an effective field theory limits are also set on the dark matter--nucleon scattering cross-section 
as well as the mass scale \mstar\ of the unknown mediating interaction for dark matter pair production in association 
with a leptonically decaying $W$.
} 

\PreprintJournalName{
JHEP
} 

\begin{document}

\title{
Search for new particles in events with one lepton and missing transverse momentum
in $pp$ collisions at $\sqrt{s}$~=~8\, \tev\ with the ATLAS detector
}

\author{The ATLAS Collaboration}

\abstract{
This paper presents a search for new particles in events with one lepton (electron or muon) and 
missing transverse momentum using 20.3~\ifb\  of proton--proton collision data at  
$\sqrt{s}=8\tev$ recorded by the ATLAS experiment at the Large Hadron Collider.
No significant excess beyond Standard Model expectations is observed.
A \wp\ with Sequential Standard Model couplings is excluded at the 95\% confidence level for masses up to 3.24~\tev.
Excited chiral bosons (\wstar) with equivalent coupling strengths are excluded for masses up to 3.21~\tev.
In the framework of an effective field theory limits are also set on the dark matter--nucleon scattering cross-section 
as well as the mass scale \mstar\ of the unknown mediating interaction for dark matter pair production in association 
with a leptonically decaying $W$.
}

\maketitle

\section{Introduction}
\label{sec:intro}

High-energy collisions at CERN's Large Hadron Collider (LHC) provide new opportunities to search for physics beyond the Standard Model (SM). 
This paper describes such a search in events containing a lepton (electron or muon) and 
missing transverse momentum using 8~\tev\ \pp\ collision data collected with the ATLAS detector during 
2012, corresponding to a total integrated luminosity of 20.3~\ifb. 

The first new-physics scenario that is considered in this paper is the Sequential Standard Model (SSM), 
the extended gauge model of ref.~\cite{ssm}. This model proposes the existence of additional heavy gauge bosons,
of which the charged ones are commonly denoted \wp. The \wp\ has the same couplings to fermions as the SM \w~boson 
and a width that increases linearly with the \wp\ mass. The coupling of the \wp\ to $WZ$ is set to zero.
Similar searches~\cite{cdf:Wprime2010,cms:wprime2011-2,cms:wprime2013-1,atlas:wprime_2010_pub,atlas:wprime_2011-2_pub,atlas:wprime_2012_pub}
have been performed using $\sqrt{s} = 1.96$ TeV $p\bar{p}$ collision data by the CDF Collaboration, 
$\sqrt{s} = 7$ TeV $pp$ collision data by the ATLAS Collaboration as well as $\sqrt{s} = 7$ TeV and $\sqrt{s} = 8$ TeV data by the CMS Collaboration.

The second new-physics scenario that is considered originates from ref.~\cite{wzstar_motivate} and proposes 
the existence of charged partners, denoted \wstar, of the chiral boson excitations described in ref.~\cite{wzstar}.
The anomalous (mag\-netic-moment type) coupling of the \wstar\ leads to kinematic distributions significantly different 
from those of the \wp\ as demonstrated in the previous  ATLAS search~\cite{atlas:wprime_2012_pub}
that was performed using 7~\tev\ \pp\ collision data collected in 2011 corresponding to an integrated luminosity of 4.7~\ifb. 
In the analysis presented in this paper the search region is expanded to higher masses and the sensitivity is considerably improved in the region covered by the previous search.

The third new-physics scenario considered is of direct production of weakly interacting candidate dark matter (DM) particles. 
These particles can be pair-produced at the LHC, $pp\rightarrow\chi\bar{\chi}$, via a new intermediate state. 
Since DM particles do not interact with the detector material, these events can be detected if there is associated initial-state radiation of a SM particle~\cite{Birkedal:2004xn,Goodman:2010yf,Bai:2010hh,Goodman:2010ku}. 
The Tevatron and LHC collaborations have reported limits on the cross-section of $p\bar{p}/pp\rightarrow\chi\bar{\chi}+X$ 
where $X$ is a hadronic jet~\cite{cdfWIMPjet,atlasWIMPjet,cmsWIMPjet},  a photon~\cite{atlasWIMPphoton,cmsWIMPphoton}, 
a hadronically decaying \w\ or \z\ boson~\cite{atlasWIMPhadronicWZ} or a leptonically decaying \z\ boson~\cite{atlasWIMPleptonicZ}. 
Previous LHC results have also been reinterpreted to set limits on the scenario where $X$ is a leptonically decaying \w\ boson \cite{monolep}.
This analysis is the first direct ATLAS search for this case. Limits are reported for the \mbox{DM--nucleon} scattering cross-section as well as the
mass scale, \mstar, of a new SM--DM interaction expressed in an effective field theory (EFT) as a four-point contact interaction
~\cite{Beltran:2010ww,Cao:2009uw,Rajaraman:2011wf,Fox:2011pm,Cheung:2012gi,Cotta:2012nj}.
As discussed in the literature, e.g. refs.~\cite{Buchmueller:2013dya,Busoni:2013lha}, the EFT formalism is not always an appropriate approximation 
but this issue is not addressed any further in this paper. Four effective operators are used as a representative set based on the definitions in 
ref.~\cite{Goodman:2010ku}: D1 scalar, D5 vector (both constructive and destructive interference cases are considered, the former denoted by D5c and the 
latter by D5d) and D9 tensor.

The analysis presented here identifies event candidates in the electron and muon channels,
sets separate limits and then combines these assuming a common branching fraction for the two final states.
The kinematic variable used to identify the signal is the transverse mass
\begin{equation}
\mt = \sqrt{ 2 \pt \met (1 - \cos \varphi_{\ell\nu})},
\end{equation}
where \pt\ is the lepton transverse momentum, \met\ is the magnitude of the missing transverse momentum vector
and $\varphi_{\ell\nu}$ is the angle between the \pt\ and \met\ vectors.\footnote{
ATLAS uses a right-handed coordinate system with its origin at the nominal
interaction point in the centre of the detector and the $z$-axis along the
beam pipe. Cylindrical coordinates $(r,\varphi)$ are used in the transverse
plane, $\varphi$ being the azimuthal angle around the beam pipe. The pseudorapidity $\eta$ is
defined in terms of the polar angle $\theta$ by $\eta=-\ln\tan(\theta/2)$.
}

The main background to the $W'$, $W^{*}$ and DM signals comes from the tail of the \mt\ distribution from SM \w~boson production with 
decays to the same final state.
Other relevant backgrounds are \z\ boson production with decays into two leptons where one lepton is not reconstructed,
\w\ or \z\ production with decays to $\tau$ leptons where a $\tau$ subsequently decays to either an electron or a muon, and
diboson production. These are collectively referred to as the electroweak (EW) background.
There is also a contribution to the background from \ttbar\ and single-top production, collectively referred to as the top background, 
which is most important for the lowest \wps\ masses considered here, where it constitutes about 10\% of the background after event selection in the 
electron channel and 15\% in the muon channel.
Other relevant strong-interaction background sources occur when a light or heavy hadron decays semileptonically or 
when a jet is misidentified as an electron or  muon. These are referred to as the \QCDBg\ in this paper.

\FloatBarrier

\section{The ATLAS detector} 
\label{sec:det}

The ATLAS detector~\cite{atlas:detector} is a multi-purpose particle
physics detector with a forward-backward symmetric cylindrical
geometry and nearly 4$\pi$ coverage in solid angle. 
The ATLAS detector has three major components: the inner tracking detector (ID),
the calorimeter and the muon spectrometer (MS). Tracks and vertices  of charged particles are reconstructed with 
silicon pixel and silicon microstrip detectors covering $|\eta|<2.5$ and straw-tube transition radiation detectors covering $|\eta| < 2.0$,
all immersed in a homogeneous 2~T magnetic field provided by a superconducting solenoid.
The ID is surrounded by a hermetic calorimeter that covers $|\eta| <$ 4.9 and provides three-dimensional reconstruction of particle showers. 
The electromagnetic calorimeter is a liquid argon (LAr) sampling calorimeter, which uses lead absorbers for \mbox{$|\eta| <$ 3.2} 
and copper absorbers in the very forward region. The hadronic sampling calorimeter 
uses plastic scintillator tiles as the active material and iron absorbers in the region $|\eta| <$ 1.7. In the region 
1.5 $ < |\eta| < $ 4.9, liquid argon is used as the active material, with copper and/or tungsten absorbers. 
The MS surrounds the calorimeters and consists of three large superconducting toroid systems (each with eight coils) 
together with multiple layers of trigger chambers up to $|\eta|<$ 2.4 and tracking chambers, providing precision track measurements, up to $|\eta|<$ 2.7. 

\FloatBarrier

\section{Trigger and reconstruction}
\label{sec:trigAndReco}

The data used in the electron channel were recorded with a trigger requiring the presence of an energy cluster in the EM compartment of the calorimeter
(EM cluster) with $\et > 120\gev$.
For the muon channel, matching tracks in the MS and ID with combined $\pt > 36\gev$ are
used to select events. In order to compensate for the small loss in the selection efficiency at high \pt\ due to this matching, 
events are also recorded if a muon with $\pt>40\gev$ and $|\eta|<1.05$ is found in the MS.
The average trigger efficiency (measured with respect to reconstructed objects) is above 99\% in the electron channel and 
80\%--90\% in the muon channel for the region of interest in this analysis.

Each EM cluster with \mbox{$\et > 125\gev$} and $|\eta| < 1.37$ or $1.52 < |\eta| < 2.47$
is considered as an electron candidate if it is matched to an ID track. 
The region $1.37 \leq |\eta| \leq 1.52$ exhibits degraded energy resolution due to the transition from 
the central region to the forward regions of the calorimeters and is therefore excluded.
The track and the cluster must satisfy a set of identification criteria
that are optimised for the conditions of many proton--proton collisions in the same or nearby beam bunch crossings 
(in-time or out-of-time pile-up, respectively)~\cite{atlas:egamreco}.  These criteria require the shower profiles to be consistent with those expected 
for electrons and impose a minimum requirement on the amount of transition radiation that is present. In addition, 
to suppress background from photon conversions, a hit in the first layer of the pixel detector is required if an active pixel sensor is traversed. 
The electron's energy is obtained from the calorimeter measurements while its direction is obtained from the associated track.
In the high-\et\ range relevant for this analysis, the electromagnetic calorimeter energy resolution is measured in data to be
1.2\% in the central region and 1.8\% in the forward region~\cite{atlas:egamma_perf}.
These requirements result in about a 90\% identification efficiency for electrons with \mbox{$\et > 125 \gev$}.

Muons are required to have a $\pt > 45\gev$, where the momentum of the muon is obtained by combining the ID and MS measurements. 
To ensure an accurate measurement of the momentum, muons are required to have hits in three MS layers 
and are restricted to the ranges $|\eta| < 1.0$ and $1.3 < |\eta| < 2.0$.  
Some of the chambers in the region $1.0 < |\eta| < 1.3$ were not yet installed, hence the momentum resolution 
of MS tracks is degraded in this region.  Including the muon candidates with an \eta-range 2.0 $<|\eta|<$ 2.5 would lead to an increase in the signal selection efficiency 
of up to 12\% for lower \wp\ masses and of up to 3\% for a \wp\ mass of 3 TeV.
However, the background levels in the signal region would increase by more than 15\%.
Therefore, the previously stated $\eta$ restrictions are retained. 
For the final selection of good muon candidates, the individual ID and MS momentum measurements are required to be 
in agreement within 5 standard deviations. 
The average momentum resolution is about 15\%--20\% at $\pt = 1\tev$.
About 80\% of the muons in the \eta-range considered are reconstructed, with most of the loss coming from regions without 
three MS layers. 

The \met\ in each event is evaluated by summing over energy-calibrated physics objects
(jets, photons and leptons) and adding corrections for calorimeter deposits not associated with these objects~\cite{atlas:met2011}.

This analysis makes use of all of the $\sqrt{s} = 8\tev$ data collected in 2012  for which the relevant detector systems were operating properly 
and all data quality requirements were satisfied. The integrated luminosity of the data used in this study is 20.3~\ifb\ for both the electron 
and muon decay channels.
The uncertainty on this measurement is 2.8\%, which is derived following the methodology detailed in ref.~\cite{atlas:lumi7Tev}.

\FloatBarrier

\section{Monte Carlo simulation}
\label{sec:mcSim}

With the exception of the \QCDBg, which is estimated from data, expected signals and backgrounds are 
evaluated using simulated Monte Carlo samples and normalised using the calculated cross-sections and the integrated luminosity of the data.

The \wp\ signal events are generated at leading order (LO) with \pythia\ v8.165~\cite{pythia,pythia8} using the MSTW2008 LO~\cite{mstw}
parton distribution functions (PDFs).  \pythia\ is also used for the fragmentation and hadronisation  of \wsl\ events that are 
generated at LO with \calchep\ v3.3.6~\cite{calchep} using the CTEQ6L1 PDFs~\cite{cteq6l}. 
DM signal samples are generated at LO with \madgraphFive\ v1.4.5~\cite{madgraph} using the MSTW2008 LO PDFs, interfaced to \pythia\ v8.165.

The \wzbg\ boson and \ttbar\ backgrounds are generated at next-to-leading order (NLO) with \powhegbox\ r1556 \cite{powheg} 
using the CT10 NLO~\cite{CT10} PDFs. For the  \wzbg\  backgrounds, fragmentation and hadronisation is performed with  \pythia\ v8.165,
while for  \ttbar\  \pythia\ v6.426 is used. The single-top background is generated at NLO with \mcatnlo\ v4.06~\cite{mcatnlo} using the CT10 NLO PDFs 
for the $Wt$- and $s$-channels, and with \acerMC\ v3.8~\cite{acermc} using the CTEQ6L1 PDFs for the $t$-channel. 
Fragmentation and hadronisation for the \mcatnlo\ samples are performed with \herwig\ v6.520~\cite{herwig}, using 
\jimmy\ v4.31~\cite{jimmy} for the underlying event, whereas \pythia\ v6.426 is used for the \acerMC\ samples.
The $WW$, $WZ$ and $ZZ$ diboson backgrounds are generated at LO with \sherpa\ v1.4.1~\cite{sherpa} using the CT10 NLO PDFs.

The \pythia\ signal model for \wp\ has \vminusa\ SM couplings to fermions but does not include
interference between the \w\ and \wp. For both \wp\ and \wstar, decay channels beside $e\nu$ and $\mu\nu$,
notably $\tau\nu$, $ud$, $sc$ and $tb$, are included in the calculation of the widths but are not explicitly included as signal or background.
At high mass ($\mwp > 1~\tev$), the total width is about 3.5 \% of the pole mass, and
the branching fraction to each of the lepton decay channels is 8.2\%.

For all samples, final-state photon radiation from leptons is handled by 
\photos~\cite{photos}. The ATLAS full detector simulation~\cite{atlas:sim} based on \geant4~\cite{geant} is used to
propagate the particles and account for the response of the detector.
For the underlying event, the ATLAS tune AUET2B~\cite{mc11tune}  is used for \pythia\  6 and AU2~\cite{mc11tune1} is used for  \pythia\  8, while
AUET2~\cite{mc11tune2} is used for the \herwig\ with \jimmy.
The effect of pile-up  is incorporated into the simulation by overlaying additional minimum-bias 
events generated with \pythia\ onto the  generated hard-scatter events. Simulated events are weighted to match the 
distribution of the number of interactions per bunch crossing observed in data, but are otherwise 
reconstructed in the same manner as data.

The \wlnu\ and \zll\ cross-sections are calculated at next-to-next-to-leading order (NNLO) in QCD with ZWPROD~\cite{zwprod}
using MSTW2008 NNLO PDFs. Consistent results are obtained using VRAP v0.9~\cite{vrap} and FEWZ v3.1b2~\cite{fewz,fewz2}.
Higher-order electroweak corrections are calculated with MCSANC~\cite{mcsanc}.
Mass-dependent $K$-factors obtained from the ratios of the calculated higher-order cross-sections
 to the cross-sections of the generated samples are used to scale $W^{+}$, $W^{-}$ and $Z$ backgrounds separately.
The \wpl\ cross-sections are calculated in the same way, except that the electroweak corrections beyond 
final-state radiation are not included because the calculation for the SM \w\ cannot be applied directly. 
Cross sections for \wsl\ are kept at LO due to the non-renormalisability of the model at higher orders in QCD. 
The \ttbar\ cross-section is also calculated at NNLO including resummation of next-to-next-to-leading logarithmic (NNLL) soft gluon terms 
obtained with \topPP\ v2.0~\cite{Cacciari:2011hy,Baernreuther:2012ws,Czakon:2012zr,Czakon:2012pz,Czakon:2013goa,Czakon:2011xx} 
for a top quark mass of $172.5 \gev$.
The $W'$, $W^{*}$,  and DM particle signal cross-sections are listed in tables~\ref{tab:xsec_sig} and~\ref{tab:xsec_sigdm}.
The most important background cross-sections are listed in table~\ref{tab:xsec_bg}.

\begin{table}[!htbp]
\caption{Predicted values of the cross-section times branching fraction (\xbr) for \wpl\ and  \wsl. 
The \xbr\ for \wpl\ are at NNLO while those for \wsl\ are at LO.  
The values are given per channel, with $\ell=e$ or $\mu$.
}
\label{tab:xsec_sig}
\begin{center}
\begin{tabular}{c|l|l}
\hline
\hline
Mass   & \multicolumn{1}{c|}{\wpl}      & \multicolumn{1}{c}{\wsl}  \\ \relax  
[\gev] & \multicolumn{1}{c|}{\xbr\ [pb]} & \multicolumn{1}{c}{\xbr\ [pb]}   \\
\hline
\phantom{0}300 & \phantom{}149.0       &   \\ 
\phantom{0}400 & \phantom{0}50.2       & \phantom{0}37.6        \\ 
\phantom{0}500 & \phantom{0}21.4       & \phantom{0}16.2        \\ 
\phantom{0}600 & \phantom{0}10.4       & \phantom{00}7.95        \\ 
\phantom{0}750 & \phantom{00}4.16     & \phantom{00}3.17         \\ 
\phantom{}1000 & \phantom{00}1.16     & \phantom{00}0.882       \\ 
\phantom{}1250 & \phantom{00}0.389   & \phantom{00}0.294         \\ 
\phantom{}1500 & \phantom{00}0.146    & \phantom{00}0.108        \\ 
\phantom{}1750 & \phantom{00}0.0581  & \phantom{00}0.0423      \\ 
\phantom{}2000 & \phantom{00}0.0244  & \phantom{00}0.0171       \\ 
\phantom{}2250 & \phantom{00}0.0108  \phantom{0}  & \phantom{00}0.00700  \\ 
\phantom{}2500 & \phantom{00}0.00509 \phantom{0} & \phantom{00}0.00290  \\
\phantom{}2750 & \phantom{00}0.00258 &  \phantom{00}0.00120  \\ 
\phantom{}3000 & \phantom{00}0.00144 &  \phantom{00}4.9$\times 10^{-4}$   \\ 
\phantom{}3250 & \phantom{00}8.9$\times 10^{-4}$ & \phantom{00}2.0$\times 10^{-4}$  \\ 
\phantom{}3500 & \phantom{00}5.9$\times 10^{-4}$ & \phantom{00}8.0$\times 10^{-5}$   \\ 
\phantom{}3750 & \phantom{00}4.2$\times 10^{-4}$ &  \phantom{00}3.2$\times 10^{-5}$   \\ 
\phantom{}4000 & \phantom{00}3.1$\times 10^{-4}$ &  \phantom{00}1.3$\times 10^{-5}$   \\  
\hline
\end{tabular}
\end{center}
\end{table}

\begin{table}[!htbp]
\caption{
Predicted values of \xbr\ for DM signal with different mass values, $m_{\chi}$.  
The values of \mstar\ used in the calculation for a given operator are also shown. 
The cross-sections are at LO, and the values are given for the sum of three lepton flavours $\ell=e,\mu,\tau$.
}
\label{tab:xsec_sigdm}
\begin{center}
\begin{tabular}{c|l|l|l|l}
\hline
\hline
    & \multicolumn{4}{c}{DM production}  \\ 
$m_{\chi}$  & \multicolumn{4}{c}{$\xbr$ [pb]}       \\ \relax
[\gev]   &  \multicolumn{1}{c}{D1} & \multicolumn{1}{c}{D5d} & \multicolumn{1}{c}{D5c} & \multicolumn{1}{c}{D9} 
 \\
             &  \multicolumn{1}{c}{\mstar$\ = 10 \GeV$}  & \multicolumn{1}{c}{\mstar$\ = 100 \GeV$} & \multicolumn{1}{c}{\mstar$\ = 1 \TeV$} & \multicolumn{1}{c}{\mstar$\ = 1 \TeV$} 
 \\

 \hline
\phantom{000}1     & 439      & 72.2   & 0.0608   & 0.0966    \\
\phantom{0}100    & 332    &  70.8  &  0.0575  &  0.0870   \\
\phantom{0}200    & 201    &  58.8  &  0.0488  & 0.0695   \\
\phantom{0}400 & \phantom{0}64.6   & 32.9   & 0.0279   & 0.0365   \\
\phantom{}1000 & \phantom{00}1.60  & \phantom{0}2.37   &  0.00192  &  0.00227  \\
\phantom{}1300  & \phantom{00}0.213 & \phantom{0}0.454  &  0.000351 &  0.000412 \\
\hline
\hline
\end{tabular}
\end{center}
\end{table}

\begin{table}[!ht]
\caption{
Predicted values of \xbr\ for the leading backgrounds.
The value for $\ttbar\rightarrow\ell X$ includes all final states with at least one
lepton ($e$, $\mu$ or $\tau$).
The others are exclusive and are used for both $\ell=e$ and $\ell=\mu$.
All cross-sections are at NNLO.
}
\label{tab:xsec_bg}
\begin{center}
\begin{tabular}{l|l}
\hline
\hline
Process & $\xbr$ [pb] \\
\hline
\wlnu\ & \phantom{}12190\phantom{.0} \\
\zgll\  ($m_{\zg}>60 \gev$)  & \phantom{0}1120\phantom{.0} \\
 \ttbarl &   \phantom{00}137.3 \\
\hline
\hline
\end{tabular}
\end{center}
\end{table}

Uncertainties on the \wp\ cross-section and the \wzbg\ background cross-sections are estimated from variations of the renormalisation and factorisation
scales, PDF$+\alpha_{\mathrm{s}}$ variations and PDF choice.
The scale uncertainties are estimated by varying both the renormalisation and factorisation 
scales simultaneously up or down by a factor of two. The resulting maximum variation from the two fluctuations is taken as the symmetric scale uncertainty.
The PDF$+\alpha_{\mathrm{s}}$ uncertainty is evaluated using 90\% confidence level (CL) eigenvector and 90\% CL $\alpha_{\mathrm{s}}$ variations of the 
nominal MSTW2008 NNLO PDF set and combined with the scale uncertainty in quadrature.
The PDF choice uncertainty is evaluated by comparing the central values of the MSTW2008 NNLO, CT10 NNLO, NNPDF 2.3 NNLO~\cite{nnpdf}, 
ABM11 5N NNLO~\cite{abm11} and HERAPDF 1.5 NNLO~\cite{Radescu:2011cn} PDF sets.
The envelope of the PDF central value comparisons and the combination of the scale and PDF$+\alpha_{\mathrm{s}}$ 
uncertainties is taken as the total uncertainty on the differential cross-section as a function of the invariant mass of the lepton--neutrino system ($m_{\ell\nu}$).
The PDF and $\alpha_{\mathrm{s}}$ uncertainties on the $t\bar{t}$ cross-section are calculated using the PDF4LHC prescription~\cite{pdf4lhc} 
with the MSTW2008 68\% CL NNLO, CT10 NNLO and NNPDF2.3 5f FFN PDF error sets added in quadrature to the scale uncertainty. 
The systematic uncertainty arising from the variation of the top mass by $\pm$1~GeV is also added in quadrature.

An additional uncertainty on the differential cross-section due to the beam energy uncertainty is calculated as function of $m_{\ell\nu}$
for the charged-current Drell--Yan process with VRAP at NNLO using  CT10 NNLO PDFs  by taking a 0.66\% uncertainty on the 
energy of each 4 TeV  proton beam as determined in ref. \cite{Wenninger:1546734}.  
The size of this uncertainty is observed to be about 2\% (6\%) at $m_{\ell\nu} =$ 2 (3) TeV. 
The calculated uncertainties are propagated to both the $W$ and \wps\ processes in order to derive uncertainties on the background levels as well as the signal selection efficiencies in each signal region. 

Uncertainties are not reported on the cross-sections for the \wstar\ due to the breakdown of higher-order corrections for non-renormalisable models.
However, uncertainties on the signal selection efficiency for the \wstar\ are evaluated using the same relative differential cross-section uncertainty as for the \wp.
Uncertainties on DM production are evaluated using 68\% confidence level eigenvector variations of the nominal MSTW2008 LO PDF set 
as in~\cite{atlasWIMPhadronicWZ}.

\FloatBarrier

\section{Event selection}
\label{sec:eventSel}

The primary vertex for each event is required to have at least three tracks with $\pt > 0.4$~\gev\ and 
to have a longitudinal distance less than 200~mm from the centre of the collision region.
There are on average 20.7 interactions per event in the data used for this analysis.
The primary vertex is defined to be the one with the highest summed track $\pt^2$.
Spurious tails in the \met\ distribution, arising from calorimeter noise and other detector problems are suppressed by checking the quality
of each reconstructed jet and discarding events containing reconstructed jets of poor quality, following the description given in ref.~\cite{atlas:2012:jet_cleaning}.
In addition, the ID track associated with the electron or muon is required to be compatible
with originating from the primary vertex by requiring that the transverse distance of closest approach, $d_0$, satisfies $|d_0|<1~(0.2)$~mm
and  longitudinal distance, $z_0$, satisfies $|z_0|<5~(1)$~mm for the electron (muon).
Events are required to have exactly one electron candidate with  \mbox{$\et > 125\gev$}  or one muon candidate with  $\pt > 45\gev$ satisfying
these requirements and the identification criteria described in section~\ref{sec:trigAndReco}. 
In the electron channel, events having additional electrons with  $\et > 20\gev$, 
passing all electron identification criteria, are discarded.  Similarly, in the muon channel, 
events having additional muon candidates with a \pt\ threshold of 20 GeV are discarded. 
 
To suppress the \QCDBg, the lepton is required to be isolated. In the electron channel,
the isolation energy is measured with the calorimeter in a cone \mbox{$\Delta R = \sqrt{(\Delta \eta)^2 + (\Delta \varphi)^2} = 0.2$}
around the electron track, and the requirement is $\Sigma E^{\mathrm{calo}}_{\mathrm{T}}< 0.007\times\et + 5 \gev$,
where the sum includes all calorimeter energy clusters in the cone excluding those that are attributed to the electron. 
The scaling of the isolation requirement with the electron \et\ reduces the efficiency loss due to radiation from the electron at high \et.
In the muon channel, the isolation energy is measured using ID tracks with $\pttrack > 1$~\gev\ in a cone $\Delta R = 0.3$ around the muon track.
The isolation requirement is \mbox{$\sumpttrack < 0.05 \times \pt$}, where the muon track is excluded from the sum.
As in the electron channel, the scaling of the isolation requirement with the muon \pt\ reduces the efficiency loss due to radiation from the muon at high \pt.

An \met\ requirement is imposed to select signal events and to further suppress the contributions from the multi-jet and SM $W$ backgrounds.
In both channels, the requirement placed on the charged lepton
\pt\ is also applied to the \met: $\met > 125 \gev$ for the electron channel and $\met > 45 \gev$ for the muon channel.

The \QCDBg\ around the Jacobian peak of the \mt\ distribution is evaluated using the \emph{matrix method} as described in ref.~\cite{atlas:ttbar2010} in both the electron and muon channels. The high-mass tail of the distribution is then fitted by a power-law function in order to determine the level of the \QCDBg\ 
in the region used to search for new physics. 
In the electron channel, the \QCDBg\ constitutes about 2\%--4\% of the total background at high \mt. Consistent results are obtained using the 
\emph{inverted isolation} technique described in ref.~\cite{atlas:wprime_2010_pub}.
In the muon channel, the \QCDBg\ constitutes about 1\%--3\% of the total background at high \mt.
The uncertainty of the \QCDBg\ is determined by varying the selection requirements used to define the control region and 
by varying the \mt\ threshold of the fitting range used in the extrapolation to high \mt.

The same reconstruction criteria and event selection are applied to both the data and simulated samples.
Figure~\ref{fig:final_mt} shows the \pt, \met, and \mt\ spectra for each channel after event selection
for the data, the expected background and three examples of \wp\ signals at different masses.
Prior to investigating if there is evidence for a signal, the agreement between the data and the predicted background 
is established for events with \mt\ $<\ $252~\gev, the lowest \mt\ threshold used to search for new physics.
The optimisation of the \mt\ thresholds for event selection is described below.
The agreement between the data and expected background is good.
Table~\ref{tab:nbg} shows an example of how different sources contribute to the background
for $\mt>1500 \gev$, the region used to search for a \wp\ with a mass of 2000~\gev. 
The  \wlnu\ background is the dominant contribution for both the electron and muon channels.
The \zll\ background in the electron channel is smaller than in the muon channel due to calorimeters 
having larger $\eta$ coverage than the MS, and the electron energy resolution being better than the muon
momentum resolution at high \pt.

\begin{table}[!htbp]
\caption{
Expected numbers of events from the various background
sources in each decay channel
for \mbox{$\mt>1500 \gev$}, the region used to
search for a \wp\ with a mass of 2000~\gev.
The \wlnu\ and \zll\ rows include the expected contributions from the $\tau$-lepton.
The uncertainties are statistical.
}
\label{tab:nbg}
\begin{center}
\begin{tabular}{l | l|l}
\hline
\hline
                & \multicolumn{1}{c|}{$e\nu$} & \multicolumn{1}{c}{$\mu\nu$} \\
\hline
\tspace
\wlnu          & 2.65\phantom{000} $\pm$ 0.10             &  2.28\phantom{000}  $\pm$ 0.21 \\
\zll               & 0.00163\phantom{} $\pm$ 0.00022     &  0.232\phantom{00}  $\pm$ 0.005 \\
Diboson     & 0.27\phantom{000} $\pm$ 0.23             &  0.46\phantom{000}  $\pm$ 0.23 \\
Top              & 0.0056\phantom{0} $\pm$ 0.0009        &  0.0017 \phantom{0}$\pm$ 0.0001 \\
Multi-jet            &  0.066\phantom{00} $\pm$  0.020        &  0.046 \phantom{00}$\pm$ 0.039 \\
\hline
\tspace
Total           & 2.99\phantom{000} $\pm$ 0.25 & 3.01\phantom{000} $\pm$ 0.31 \\
\hline
\hline
\end{tabular}
\end{center}
\end{table}

\begin{figure*}[!htbp]
  \centering
  \includegraphics[width=0.49\textwidth]{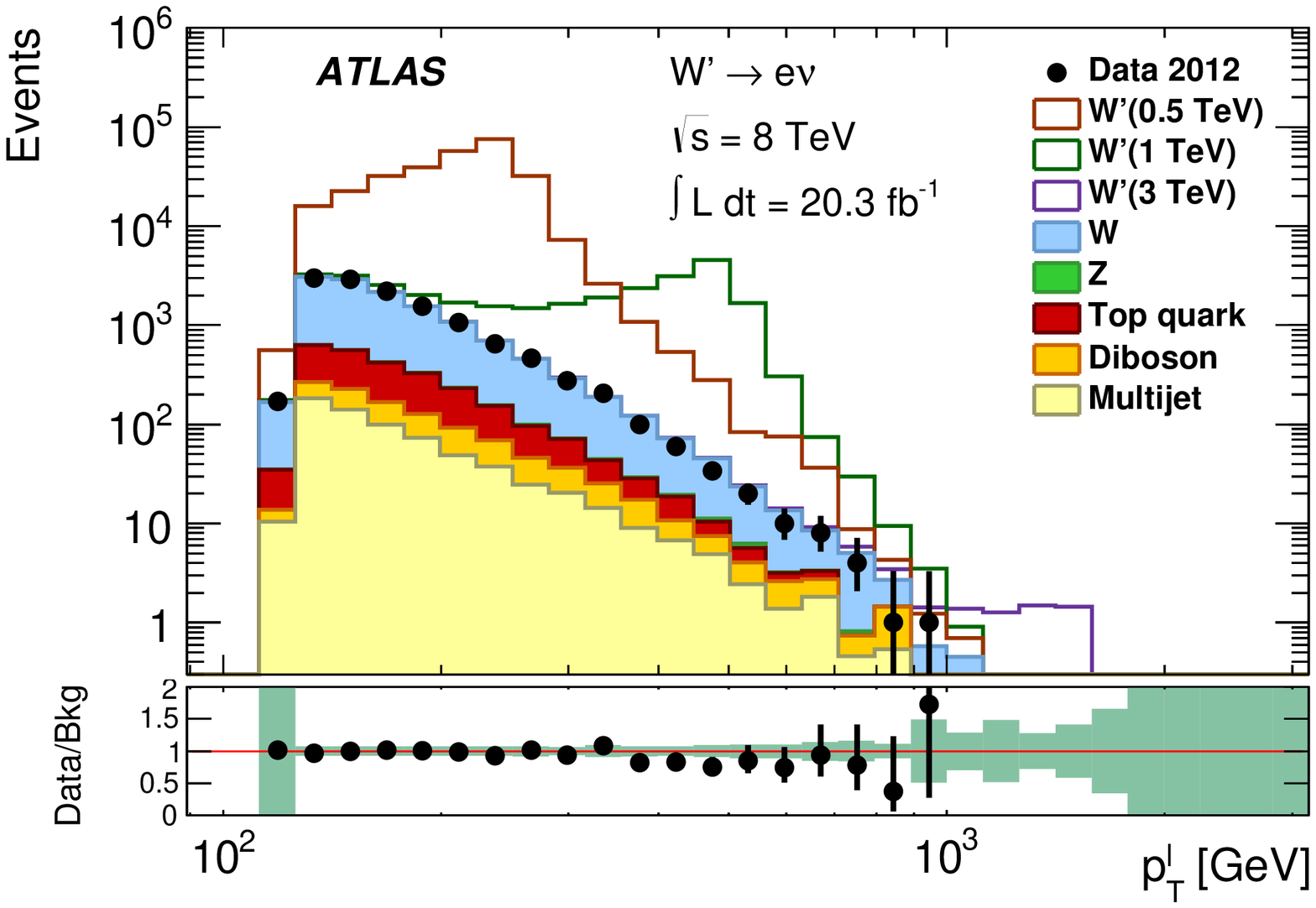}
  \includegraphics[width=0.49\textwidth]{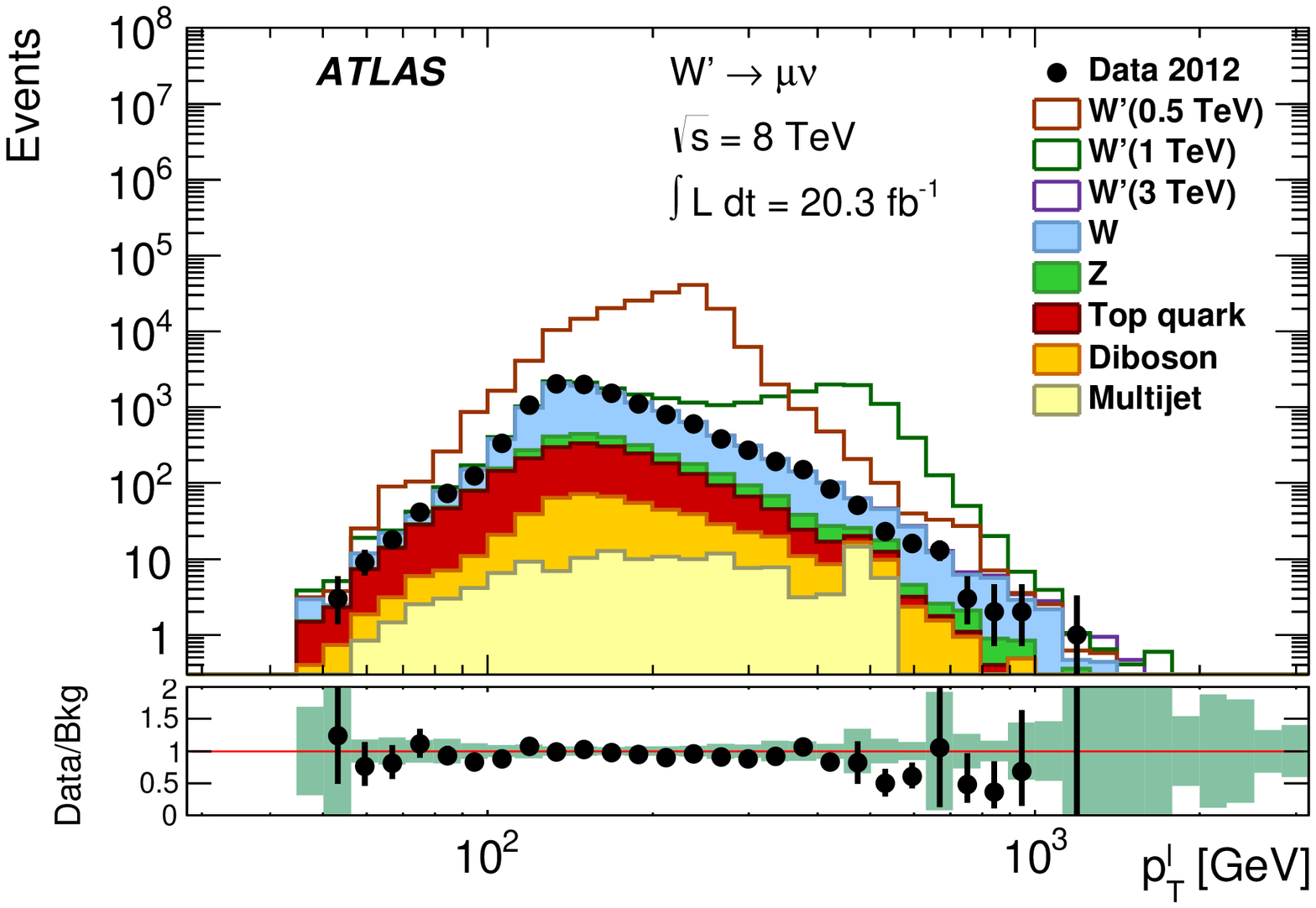}
  \includegraphics[width=0.49\textwidth]{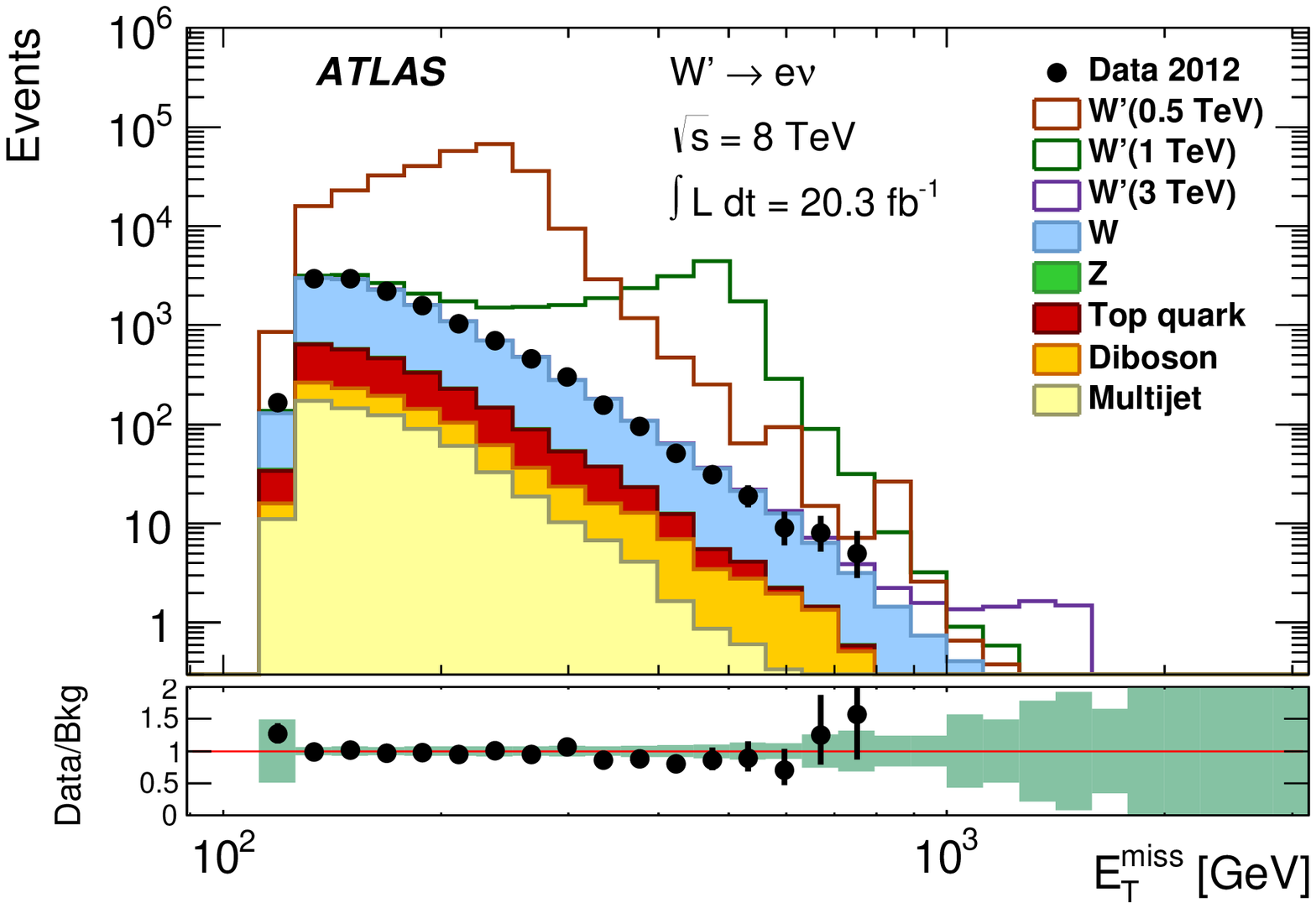}
  \includegraphics[width=0.49\textwidth]{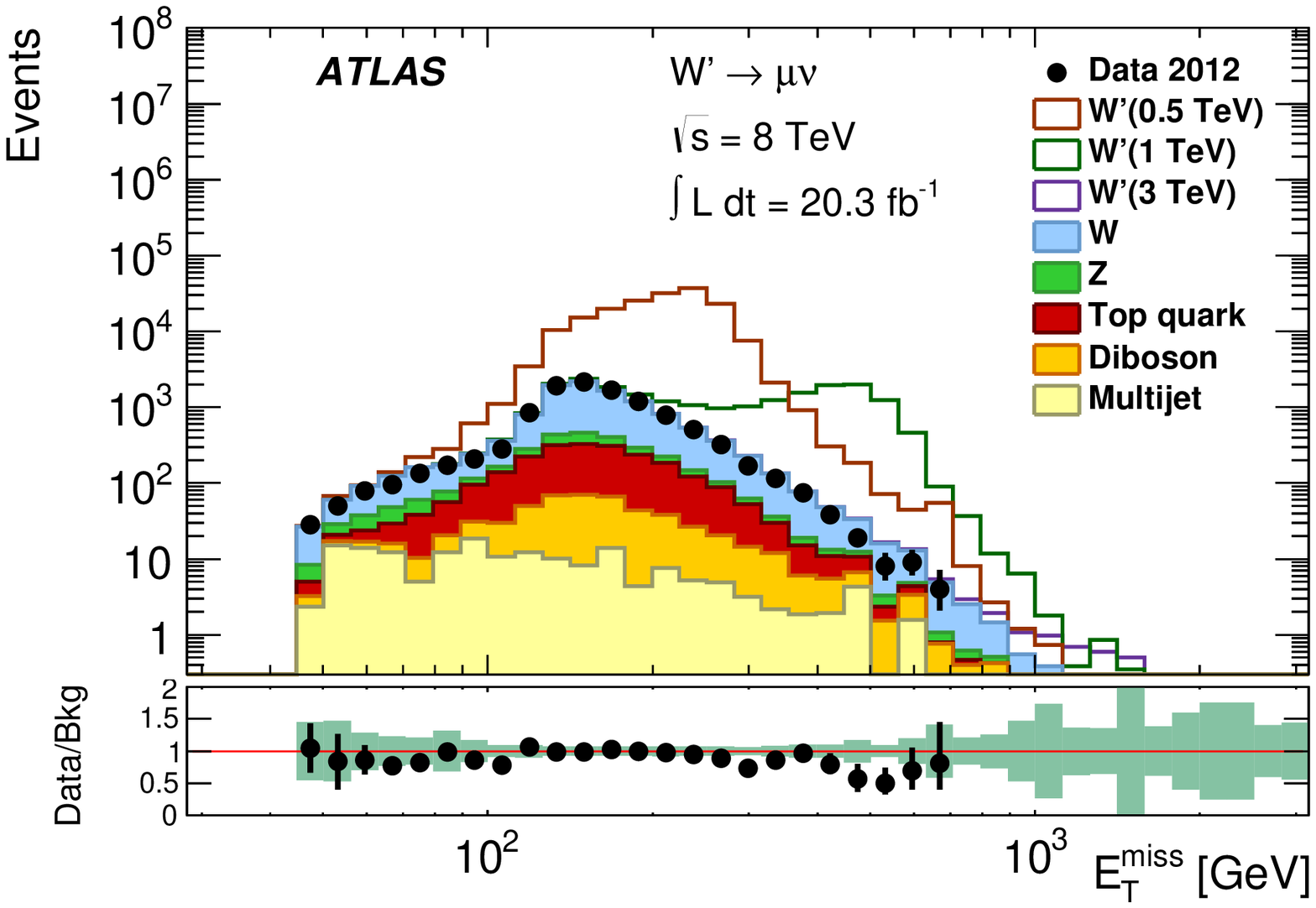}
  \includegraphics[width=0.49\textwidth]{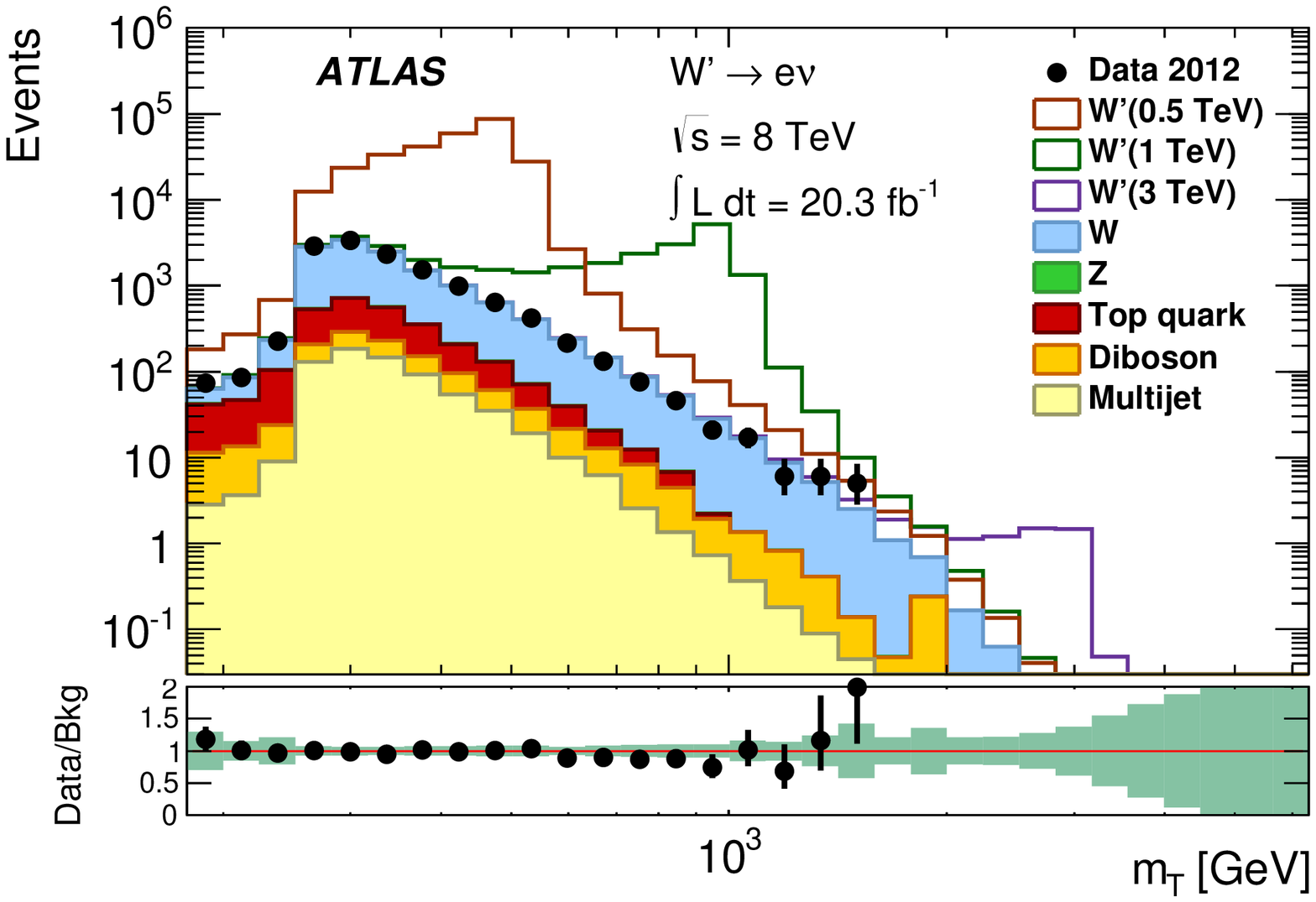}
  \includegraphics[width=0.49\textwidth]{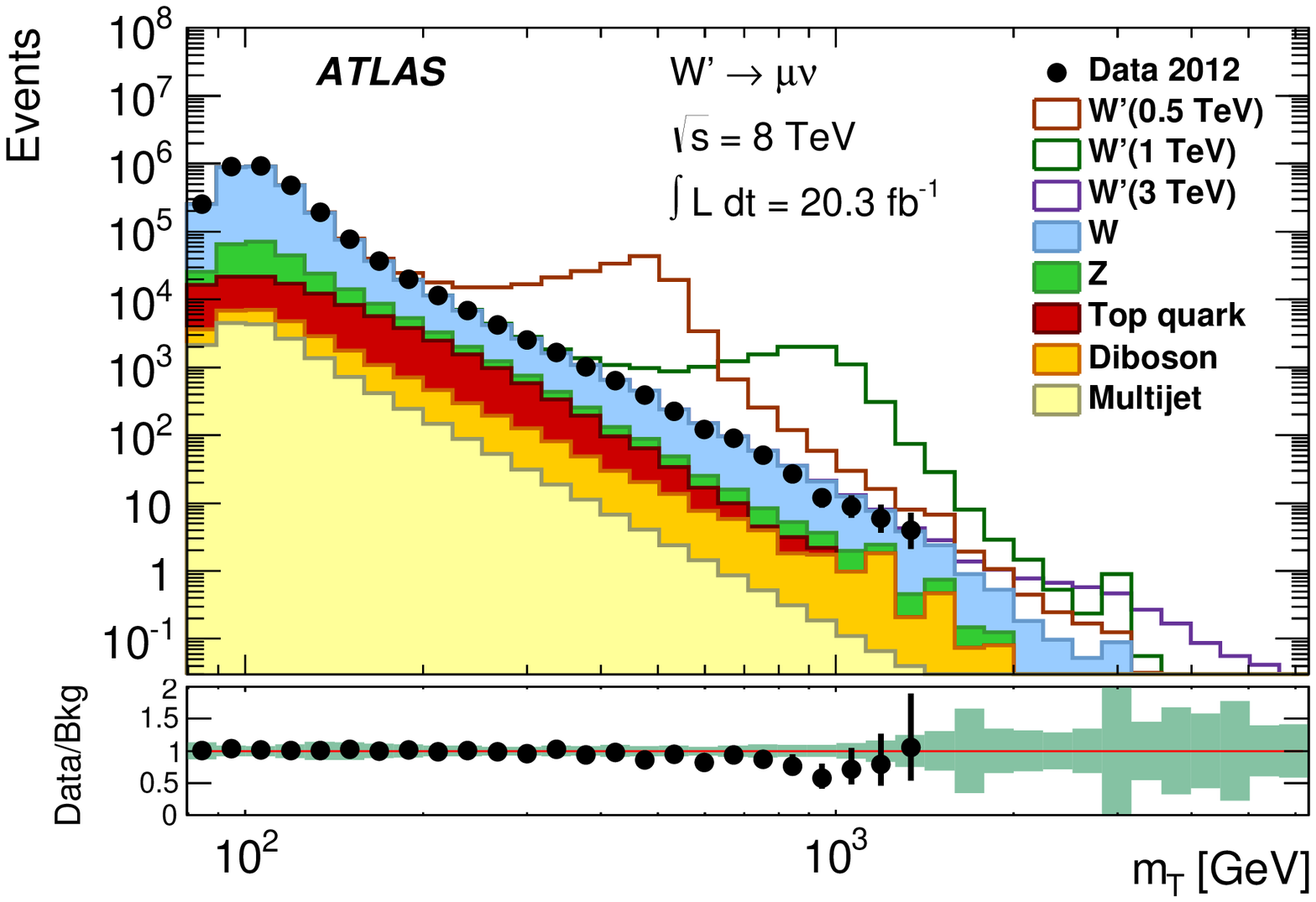}
  \caption{Spectra of lepton \pt\ (top), \met\ (centre) and \mt\ (bottom)
  for the electron (left) and muon (right) channels after the event selection. 
  The spectra of \pt\  and \met\ are shown with the requirement \mt\ $>$ 252 \gev.
  The points represent data and the filled, stacked histograms show the predicted backgrounds.
  Open histograms are \wpl\ signals added to the background with their masses in \gev\ indicated in parentheses in the legend. 
  The signal and background samples are normalised using the integrated luminosity of the data
  and the NNLO cross-sections listed in tables~\ref{tab:xsec_sig} and~\ref{tab:xsec_bg},
  except for the \QCDBg\ which is estimated from data. The error bars on the data points
  are statistical. The ratio of the data to the total background prediction is shown below each of the distributions.
  The bands represent the systematic uncertainties on the background including the ones arising from the statistical uncertainty of
  the simulated samples.
  \label{fig:final_mt}
  }
\end{figure*}

\FloatBarrier

\section{Statistical analysis and systematic uncertainties}
\label{sec:statAnaAndSys}

A Bayesian analysis is performed to set limits on the studied processes.
For each candidate mass and decay channel, events are counted above an \mt\ threshold. 
The optimisation of \mtmin\ is done separately for \wpl\ and \wsl.
For each candidate mass, the \mtmin\ values that minimise the expected cross-section limits are obtained in the electron and muon channels separately, 
but for simplicity the lower value is used in both channels since this has a negligible impact on the final results.
A similar optimisation is performed when setting the limits on DM production, and in this case a single \mtmin\ is chosen for each operator.
The expected number of events in each channel is
\begin{equation}
\nexp = \effsig \lint \xbr + \nbg,
\end{equation}
where \lint\ is the integrated luminosity of the data sample, \effsig\ is the signal selection efficiency 
defined as the fraction of signal events that satisfy the event selection criteria as well as $\mt > \mtmin$,  \nbg\ is the expected number of background events,
and \xbr\ is the cross-section times branching fraction. 
Using Poisson statistics, the likelihood to observe \nobs\ events is
\begin{equation}
  {\cal L}(\nobs|\xbr) = \frac{(\lint \effsig \xbr + \nbg)^{\nobs} \mathrm{e}^{-(\lint \effsig \xbr + \nbg)}}{\nobs!}.
\label{eqn:lhood}
\end{equation}
Uncertainties are included by introducing nuisance parameters \npi, each with
a probability density function \pdfi, and integrating the product of
the Poisson likelihood with the probability density function.
The integrated likelihood is
\small
\begin{equation}
  {\cal L}_B(\nobs|\xbr) =  \int  {\cal L}(\nobs|\xbr)   \prod \pdfi d\npi,
\label{eqn:lhoodn}
\end{equation}
\normalsize
where a log-normal distribution is used for the \pdfi.
The nuisance parameters are taken to be: \lint, \effsig\ and \nbg, with the appropriate correlation accounted for between the first and the third parameters.

The measurements in the two decay channels are combined assuming the same branching fraction for each.
Equation (\ref{eqn:lhoodn}) remains valid with the Poisson likelihood replaced by the product
of the Poisson likelihoods for the two channels.
The integrated luminosities for the electron and muon channels are fully correlated.
For \wpsl\  the signal selection efficiencies and background levels are partly correlated with each other and 
between the two channels due to the full correlation of the cross-section uncertainties. 
If these correlations were not included, the observed \xbr\ limits would improve by 25\%--30\% for the lowest mass points,
a few percent for the intermediate mass points and by about 10\% for the highest mass points.

Bayes' theorem gives the posterior probability that the signal has signal strength \xbr:
\begin{equation}
\ppost(\xbr|\nobs) = N {\cal L}_{B}(\nobs|\xbr) \, \pprior(\xbr)
\label{eqn:ppost}
\end{equation}
where $\pprior(\xbr)$ is the assumed prior probability, here chosen to be flat in \xbr,
for $\xbr > 0$.
The constant factor $N$ normalises the total probability to one.
The posterior probability is evaluated for each mass and decay channel as well as for their combination,
and then used to set a limit on \xbr.

The inputs for the evaluation of ${\cal L}_B$ (and hence \ppost) are \lint, \effsig, \nbg, \nobs\ and the uncertainties on
the first three. The uncertainties on \effsig\ and \nbg\ account  for experimental and theoretical systematic effects as well as the statistics of the simulated samples.
The experimental systematic uncertainties include those on the efficiencies of the electron or muon trigger, reconstruction and event/object selection.
Uncertainties in the lepton energy/momentum and \met, characterised by scale and resolution uncertainties, are also included.
Performance metrics are obtained {\it in-situ} using well-known processes such as $Z\rightarrow \ell\ell$~\cite{atlas:egamreco, Aad:2014rra, atlas:jetmet}. 
Since most of these performance metrics are measured at relatively low \pt\ their values are extrapolated to
the high-\pt\ regime relevant to this analysis using MC simulation. 
The uncertainties in these extrapolations are included but are too small to significantly affect the results.
Table~\ref{tab:syst_summary} summarises the uncertainties on the event selection efficiencies 
and the expected number of background events for the \mbox{\wpl} signal with $\mwp=2000\gev$ using $\mt>1500 \gev$, and $W^{*}$
signal with $\mws=2000\gev$ using $\mt>1337 \gev$.

\begin{table}[!htbp]
\caption{
  Relative uncertainties on the selection efficiency \effsig\ and expected number of background events \nbg\
  for a \wp\ (upper part of the table) and $W^{*}$ (lower part of the table) with a mass of 2000~\gev.
  The efficiency uncertainties include contributions from the trigger, reconstruction and event selection.
  The last row gives the total relative uncertainties.
\label{tab:syst_summary}
}
\begin{center}
\begin{tabular}{l|rr||rr}
\hline
\hline
 & \multicolumn{2}{c|}{\effsig} & \multicolumn{2}{c}{\nbg} \\
 Source                      &  \multicolumn{1}{c}{$e\nu$}  & \multicolumn{1}{c|}{$\mu\nu$}
                             &  \multicolumn{1}{c}{$e\nu$}  & \multicolumn{1}{c}{$\mu\nu$} \\
\hline
\wpl\                      &  \multicolumn{2}{c}{} \\
\hline 
 Reconstruction and  trigger efficiency                      &                  2.5\%           &                  4.1\%           &   2.7\%    &                 4.1\%  \\
Lepton energy/momentum resolution  &                   0.2\%            &                  1.4\%           &   1.9\%    &                 18\%\\
Lepton energy/momentum scale           &                  1.2\%            &                 1.8\%            &   3.5\%   &                  1.5\%\\
\met\  scale and resolution  &                                    0.1\%             &                 0.1\%          &   1.2\%   &                  0.5\%  \\
 Beam energy                &                                        0.5\%              &                 0.5\%           &    2.8\%   &                   2.1\%\\
 Multi-jet background                                    & \begin{tabular}{c}{-}\end{tabular}  & \begin{tabular}{c}{-}\end{tabular} &   2.2\%   &                    3.4\% \\
 Monte Carlo statistics                             &                  0.9\%              &                  1.3\%              &  8.5\% &                 10\%  \\
 Cross-section (shape/level)                  &                2.9\%             &                  2.8\%              &  18\%  &                 15\%  \\
\hline
 Total                                                          &                4.2\%             &                    5.6\%            &  21\% &                 27\% \\
 \hline
 \hline
\wsl\                      &  \multicolumn{2}{c}{} \\
\hline
Reconstruction and  trigger efficiency                      &                  2.7\%           &                  4.1\%           &   2.6\%    &                 4.0\%  \\
Lepton energy/momentum resolution  &                   0.4\%            &                  0.9\%           &   3.0\%    &                 17\%\\
Lepton energy/momentum scale           &                  2.4\%            &                 2.4\%            &   3.1\%   &                  1.5\%\\
\met\  scale and resolution  &                                    0.1\%             &                 0.4\%          &   3.1\%   &                  0.6\%  \\
 Beam energy                &                                        0.1\%              &                   0.1\%          &    2.5\%   &                   1.9\%\\
 Multi-jet background                                    &  \begin{tabular}{c}{-}\end{tabular}  & \begin{tabular}{c}{-}\end{tabular}   &   1.8\%   &                    2.6\% \\
 Monte Carlo statistics                             &                  1.2\%              &                  1.8\%              &  6.7\% &                 8.6\%  \\
 Cross-section (shape/level)                  &                0.2\%             &                  0.2\%              &  17\%  &                 15\%  \\
\hline
 Total                                                          &                3.9\%             &                    5.1\%            &  19\% &                 25\% \\
\hline
\hline
\end{tabular}
\end{center}
\end{table}

\FloatBarrier

\section{Results}
\label{sec:results}

The inputs for the evaluation of ${\cal L}_B$  are listed in tables~\ref{tab:limit_input},~\ref{tab:limit_inputws} and~\ref{tab:limitdm_input}. 
The uncertainties on \effsig\ and \nbg\ account for all relevant experimental and theoretical effects except for the uncertainty on the integrated luminosity.
The latter is included separately and is correlated between signal and background.
The tables also list the predicted numbers of signal events, \nsig, with their uncertainties accounting for the uncertainties in 
both \effsig\ and the cross-section calculation.
The maximum value for the signal selection efficiency is at $\mwp = 2000\gev$.
For lower masses, the efficiency falls because the relative \mt\ threshold, $\mtmin/\mwp$,
increases in order to reduce the background level. The contribution from $W'\rightarrow \tau\nu$ with a leptonically decaying $\tau$ is neglected. 
It would increase the signal yield by 2\%--3\% for the highest masses. 
The background level is estimated for each mass by summing over all of the background sources.

\begin{table*}[!htbp]
\caption{Inputs for the \wpl\ \xbr\ limit calculations.
The first three columns are the \wp\  mass, \mt\ threshold and decay channel.
The next two are the signal selection efficiency, \effsig, and the prediction
for the number of signal events, \nsig, obtained with this efficiency.
The last two columns are the expected number of background
events, \nbg, and the number of events observed in data, \nobs.
The uncertainties on \nsig\ and \nbg\ include contributions from the uncertainties on the
cross-sections but not from that on the integrated luminosity.
\label{tab:limit_input}
}
\begin{center}
\begin{tabular}{ccc| r@{$ ~\pm~$}l r@{$~\pm~$}l r@{$~\pm~$}l r}
\hline
\hline
\mwp  & \mtmin & Channel  & \multicolumn{2}{c}{\effsig} & \multicolumn{2}{c}{\nsig} & \multicolumn{2}{c}{\nbg} & \nobs \\ \relax
[\gev]  & [\gev]   &            \\ 
\hline
\hline
\multirow{2}{*}{300} &  \multirow{2}{*}{252}
 & $e\nu$   & 0.228 & 0.009 & 688000 & 28000 & 12900 & 820 & 12717\\
 & & $\mu\nu$   & 0.184 & 0.007 & 555000 & 21000 & 11300 & 770 & 10927\\ \hline
\multirow{2}{*}{400} &  \multirow{2}{*}{336}
& $e\nu$   & 0.319 & 0.012 & 325000 & 12000 & 5280 & 360 & 5176\\
 & & $\mu\nu$   & 0.193 & 0.007 & 196000 & 7500 & 3490 & 250 & 3317\\ \hline
\multirow{2}{*}{500} &  \multirow{2}{*}{423} 
& $e\nu$  & 0.325 & 0.013 & 141000 & 5700 & 2070 & 150 & 2017\\
 & & $\mu\nu$  & 0.186 & 0.007 & 80900 & 3200 & 1370 & 100 & 1219\\ \hline
\multirow{2}{*}{600} &  \multirow{2}{*}{474} 
& $e\nu$  & 0.397 & 0.014 & 83800 & 2900 & 1260 & 96 & 1214\\
  & & $\mu\nu$  & 0.229 & 0.009 & 48200 & 1900 & 827 & 64 & 719\\ \hline
\multirow{2}{*}{750} &  \multirow{2}{*}{597}
 & $e\nu$  & 0.393 & 0.013 & 33200 & 1100 & 456 & 45 & 414\\
  & & $\mu\nu$  & 0.226 & 0.009 & 19100 & 750 & 305 & 30 & 255\\ \hline
\multirow{2}{*}{1000} &  \multirow{2}{*}{796} 
& $e\nu$  & 0.386 & 0.012 & 9080 & 290 & 116 & 15 & 101\\
  & & $\mu\nu$  & 0.219 & 0.009 & 5160 & 220 & 84 & 10 & 58\\ \hline
\multirow{2}{*}{1250} &  \multirow{2}{*}{1002} 
& $e\nu$  & 0.378 & 0.012 & 2980 & 98 & 35.3 & 5.8 & 34\\
 & & $\mu\nu$  & 0.210 & 0.009 & 1650 & 73 & 28.3 & 4.6 & 19\\ \hline
\multirow{2}{*}{1500} &  \multirow{2}{*}{1191} 
& $e\nu$  & 0.376 & 0.014 & 1110 & 40 & 13.2 & 2.5 & 14\\
 & & $\mu\nu$   & 0.206 & 0.010 & 610 & 30 & 10.9 & 2.3 & 6\\ \hline
\multirow{2}{*}{1750} &  \multirow{2}{*}{1416} 
& $e\nu$  & 0.336 & 0.013 & 396 & 16 & 4.56 & 0.92 & 5\\
 & & $\mu\nu$   & 0.182 & 0.010 & 214 & 12 & 4.3 & 1.1 & 0\\ \hline
\multirow{2}{*}{2000} &  \multirow{2}{*}{1500} 
& $e\nu$  & 0.370 & 0.015 & 183.0 & 7.7 & 2.99 & 0.61 & 3\\
 & & $\mu\nu$  & 0.198 & 0.011 & 98.0 & 5.5 & 3.01 & 0.80 & 0\\ \hline
\multirow{2}{*}{2250} &  \multirow{2}{*}{1683} 
& $e\nu$  & 0.327 & 0.015 & 71.5 & 3.3 & 1.38 & 0.33 & 0\\
  & & $\mu\nu$  & 0.173 & 0.011 & 37.9 & 2.3 & 1.44 & 0.33 & 0\\ \hline
\multirow{2}{*}{2500} &  \multirow{2}{*}{1888} 
& $e\nu$  & 0.262 & 0.018 & 27.1 & 1.8 & 0.432 & 0.091 & 0\\
  & & $\mu\nu$  & 0.140 & 0.012 & 14.4 & 1.2 & 0.61 & 0.15 & 0\\ \hline
\multirow{2}{*}{2750} &  \multirow{2}{*}{1888} 
& $e\nu$  & 0.235 & 0.024 & 12.3 & 1.3 & 0.432 & 0.091 & 0\\
 & & $\mu\nu$   & 0.127 & 0.014 & 6.64 & 0.74 & 0.61 & 0.15 & 0\\ \hline
\multirow{2}{*}{3000} &  \multirow{2}{*}{1888} 
& $e\nu$  & 0.183 & 0.029 & 5.33 & 0.86 & 0.432 & 0.091 & 0\\
  & & $\mu\nu$  & 0.100 & 0.016 & 2.93 & 0.48 & 0.61 & 0.15 & 0\\ \hline
\multirow{2}{*}{3250} &  \multirow{2}{*}{1888} 
& $e\nu$ & 0.124 & 0.033 & 2.22 & 0.59 & 0.432 & 0.091 & 0\\
 & & $\mu\nu$  & 0.069 & 0.018 & 1.24 & 0.32 & 0.61 & 0.15 & 0\\ \hline
\multirow{2}{*}{3500} &  \multirow{2}{*}{1888} 
& $e\nu$  & 0.077 & 0.031 & 0.92 & 0.36 & 0.432 & 0.091 & 0\\
 & & $\mu\nu$   & 0.044 & 0.017 & 0.52 & 0.20 & 0.61 & 0.15 & 0\\ \hline
\multirow{2}{*}{3750} &  \multirow{2}{*}{1888} 
& $e\nu$  & 0.047 & 0.024 & 0.40 & 0.21 & 0.432 & 0.091 & 0\\
 & & $\mu\nu$  & 0.028 & 0.013 & 0.24 & 0.11 & 0.61 & 0.15 & 0\\ \hline
\multirow{2}{*}{4000} &  \multirow{2}{*}{1888} 
& $e\nu$  & 0.031 & 0.018 & 0.20 & 0.11 & 0.432 & 0.091 & 0\\
  & & $\mu\nu$  & 0.019 & 0.010 & 0.121 & 0.061 & 0.61 & 0.15 & 0\\ \hline
\hline
\end{tabular}
\end{center}
\end{table*}

\begin{table*}[!htbp]
\caption{Inputs for the \wsl\ \xbr\ limit calculations.
The columns are the same as in table \ref{tab:limit_input}.
\label{tab:limit_inputws}
}
\begin{center}
\begin{tabular}{ccc| r@{$~\pm~$}l r@{$~\pm~$}l r@{$~\pm~$}l r}
\hline
\hline
 \mws  & \mtmin & Channel  & \multicolumn{2}{c}{\effsig} & \multicolumn{2}{c}{\nsig} & \multicolumn{2}{c}{\nbg} & \nobs \\ \relax
[\gev]  & [\gev]   &            \\ 
 \hline \hline
\multirow{2}{*}{400} &  \multirow{2}{*}{317} 
& $e\nu$  & 0.196 & 0.010 & 149000 & 7400 & 6630 & 440 & 6448\\
 & & $\mu\nu$   & 0.111 & 0.005 & 84900 & 3700 & 4420 & 310 & 4230\\ \hline
\multirow{2}{*}{500} &  \multirow{2}{*}{377} 
& $e\nu$  & 0.246 & 0.011 & 80900 & 3500 & 3320 & 220 & 3275\\
 & & $\mu\nu$  & 0.140 & 0.006 & 45900 & 1900 & 2210 & 160 & 2008\\ \hline
\multirow{2}{*}{600} &  \multirow{2}{*}{448} 
& $e\nu$  & 0.257 & 0.011 & 41400 & 1800 & 1630 & 120 & 1582\\
 & & $\mu\nu$   & 0.144 & 0.006 & 23200 & 960 & 1080 & 79 & 938\\ \hline
\multirow{2}{*}{750} &  \multirow{2}{*}{564} 
& $e\nu$  & 0.248 & 0.011 & 15900 & 680 & 593 & 54 & 524\\
 & & $\mu\nu$   & 0.143 & 0.006 & 9200 & 400 & 388 & 35 & 321\\ \hline
\multirow{2}{*}{1000} &  \multirow{2}{*}{710} 
& $e\nu$  & 0.302 & 0.013 & 5390 & 230 & 203 & 24 & 177\\
 & & $\mu\nu$   & 0.174 & 0.007 & 3100 & 130 & 143 & 17 & 109\\ \hline
\multirow{2}{*}{1250} &  \multirow{2}{*}{843} 
& $e\nu$  & 0.337 & 0.013 & 2010 & 79 & 86 & 12 & 79\\
 & & $\mu\nu$  & 0.191 & 0.008 & 1140 & 50 & 65.5 & 8.5 & 40\\ \hline
\multirow{2}{*}{1500} &  \multirow{2}{*}{1062} 
& $e\nu$  & 0.296 & 0.011 & 648 & 25 & 25.8 & 4.4 & 26\\
 & & $\mu\nu$   & 0.164 & 0.007 & 360 & 16 & 20.9 & 3.8 & 12\\ \hline
\multirow{2}{*}{1750} &  \multirow{2}{*}{1191} 
& $e\nu$  & 0.324 & 0.013 & 278 & 11 & 13.2 & 2.5 & 14\\
 & & $\mu\nu$   & 0.182 & 0.009 & 156.0 & 7.6 & 10.9 & 2.3 & 6\\ \hline
\multirow{2}{*}{2000} &  \multirow{2}{*}{1337} 
& $e\nu$  & 0.341 & 0.013 & 118.0 & 4.6 & 6.8 & 1.3 & 9\\
 & & $\mu\nu$   & 0.186 & 0.010 & 64.6 & 3.3 & 5.8 & 1.4 & 3\\ \hline
\multirow{2}{*}{2250} &  \multirow{2}{*}{1416} 
& $e\nu$  & 0.391 & 0.014 & 55.5 & 2.0 & 4.56 & 0.92 & 5\\
 & & $\mu\nu$   & 0.204 & 0.010 & 28.9 & 1.5 & 4.3 & 1.1 & 0\\ \hline
\multirow{2}{*}{2500} &  \multirow{2}{*}{1683} 
& $e\nu$  & 0.337 & 0.013 & 19.80 & 0.76 & 1.38 & 0.33 & 0\\
 & & $\mu\nu$   & 0.179 & 0.010 & 10.50 & 0.57 & 1.44 & 0.33 & 0\\ \hline
\multirow{2}{*}{2750} &  \multirow{2}{*}{1888} 
& $e\nu$  & 0.322 & 0.013 & 7.84 & 0.31 & 0.432 & 0.091 & 0\\
 & & $\mu\nu$   & 0.161 & 0.011 & 3.92 & 0.27 & 0.61 & 0.15 & 0\\ \hline
\multirow{2}{*}{3000} &  \multirow{2}{*}{1888} 
& $e\nu$  & 0.382 & 0.015 & 3.80 & 0.15 & 0.432 & 0.091 & 0\\
 & & $\mu\nu$  & 0.185 & 0.011 & 1.84 & 0.11 & 0.61 & 0.15 & 0\\ \hline
\multirow{2}{*}{3250} &  \multirow{2}{*}{1888} 
& $e\nu$  & 0.437 & 0.018 & 1.770 & 0.073 & 0.432 & 0.091 & 0\\
 & & $\mu\nu$  & 0.218 & 0.014 & 0.880 & 0.056 & 0.61 & 0.15 & 0\\ \hline
\multirow{2}{*}{3500} &  \multirow{2}{*}{1888} 
& $e\nu$  & 0.474 & 0.025 & 0.766 & 0.040 & 0.432 & 0.091 & 0\\
 & & $\mu\nu$   & 0.229 & 0.016 & 0.371 & 0.027 & 0.61 & 0.15 & 0\\ \hline
\multirow{2}{*}{3750} &  \multirow{2}{*}{1888} 
& $e\nu$  & 0.498 & 0.055 & 0.320 & 0.035 & 0.432 & 0.091 & 0\\
 & & $\mu\nu$   & 0.244 & 0.029 & 0.157 & 0.019 & 0.61 & 0.15 & 0\\ \hline
\multirow{2}{*}{4000} &  \multirow{2}{*}{1888} 
& $e\nu$  & 0.487 & 0.150 & 0.124 & 0.038 & 0.432 & 0.091 & 0\\
 & & $\mu\nu$   & 0.242 & 0.073 & 0.062 & 0.019 & 0.61 & 0.15 & 0\\
\hline
\hline
\end{tabular}
\end{center}
\end{table*}

\begin{table*}[!htbp]
\caption{Inputs to the limit calculations on the pair production of DM particles for the operators D1, D5d, D5c and D9.  
Expected number of signal events for each operator is calculated for a different value of  the mass scale, 
notably $M_{*}=$ 10 GeV for D1, $M_{*}=$ 100 GeV for D5d, and  \mbox{$M_{*}=$ 1 TeV}  for operators D9 and D5c.  
The columns are the same as in table \ref{tab:limit_input}.
\label{tab:limitdm_input}
}
\begin{center}
\footnotesize
\begin{tabular}{ccc| ll l l |lcc}
\hline
\hline
$m_{\chi}$ & \mtmin & Channel  & \multicolumn{2}{c}{\effsig} & \multicolumn{2}{c}{\nsig} & \multicolumn{2}{c}{\nbg} & \nobs \\
~[\gev]  & [\gev]   &            \\ 

\hline
\hline
     & & \multicolumn{8}{c}{D1 Operator}\\
    \hline
\multirow{2}{*}{1} & \multirow{12}{*}{796} 
& $e\nu$ & {$0.0294\pm0.0044$}& & {$87000\pm13000$}& & \multirow{12}{*}{$e\nu$} &\multirow{12}{*}{$116\pm15$} & \multirow{12}{*}{101}\\
& &  $\mu\nu$& {$0.0177\pm0.0023$}& & {$52500\pm7000$}& & \multirow{12}{*}{$\mu\nu$}&\multirow{12}{*}{\phantom{0}$84\pm10$}&  \multirow{12}{*}{58}\\
\multirow{2}{*}{100}  &
& $e\nu$ & {$0.0396\pm0.0052$}& & {$89000\pm12000$}&  \\
& & $\mu\nu$ & {$0.0252\pm0.0033$}& & {$56600\pm7500$}&  \\
\multirow{2}{*}{200}  &
& $e\nu$ & {$0.0484\pm0.0057$}& & {$65800\pm7700$}&  \\
& & $\mu\nu$& {$0.0293\pm0.0034$}& & {$39900\pm4600$}&  \\
\multirow{2}{*}{400}  &
& $e\nu$ & {$0.0709\pm0.0071$}& & {$30900\pm3100$}&  \\
& & $\mu\nu$& {$0.0398\pm0.0041$}& & {$17300\pm1800$}&  \\
\multirow{2}{*}{1000}  &
& $e\nu$ & {$0.0989\pm0.0100$}& & {\phantom{0}$1070\pm110$}&  \\
& & $\mu\nu$& {$0.0621\pm0.0068$}& & {\phantom{00}$673\pm73$}&  \\
\multirow{2}{*}{1300}  &
& $e\nu$ & {$0.0964\pm0.0095$}& & {\phantom{00}$138\pm14$}&  \\
& & $\mu\nu$& {$0.0522\pm0.0048$}& & {\phantom{ 0}$75.1\pm6.9$}&  \\
    \hline
    & & \multicolumn{8}{c}{D5d Operator}\\
    \hline
\multirow{2}{*}{1} & \multirow{12}{*}{597} 
& $e\nu$ & {$0.0148\pm0.0016$}& & {\phantom{0}$7230\pm800$}& & \multirow{12}{*}{$e\nu$} & \multirow{12}{*}{$456\pm45$} & \multirow{12}{*}{414} \\
& & $\mu\nu$& {$0.0080\pm0.0011$}& & {\phantom{0}$3890\pm530$}& & \multirow{12}{*}{$\mu\nu$}& \multirow{12}{*}{$305\pm30$} & \multirow{12}{*}{255}\\
\multirow{2}{*}{100} &  
& $e\nu$ & {$0.0158\pm0.0018$}& & {\phantom{0}$7580\pm850$}&   \\
&  & $\mu\nu$& {$0.0096\pm0.0012$}& & {\phantom{0}$4600\pm580$}& \\
\multirow{2}{*}{200} & 
& $e\nu$ & {$0.0147\pm0.0015$}& & {\phantom{0}$5850\pm610$}&  \\
&  & $\mu\nu$& {$0.0086\pm0.0011$}& & {\phantom{0}$3420\pm430$}&  \\
\multirow{2}{*}{400} & 
 & $e\nu$ & {$0.0190\pm0.0020$}& & {\phantom{0}$4220\pm440$}&  \\
&  & $\mu\nu$& {$0.0113\pm0.0013$}& & {\phantom{0}$2500\pm300$}&  \\
\multirow{2}{*}{1000} & 
 & $e\nu$ & {$0.0281\pm0.0025$}& & {\phantom{00}$450\pm41$}&  \\
&  & $\mu\nu$& {$0.0177\pm0.0019$}& & {\phantom{00}$283\pm30$}&  \\
\multirow{2}{*}{1300} & 
 & $e\nu$ & {$0.0291\pm0.0028$}& & {\phantom{ 0}$89.3\pm8.5$}&  \\
& & $\mu\nu$& {$0.0167\pm0.0018$}& & {\phantom{ 0}$51.1\pm5.4$}&  \\
    \hline
    & & \multicolumn{8}{c}{D5c Operator}\\
    \hline
\multirow{2}{*}{1} & \multirow{12}{*}{843}
 & $e\nu$ & {$0.0737\pm0.0047$} & & {$\phantom{0}30.3\pm1.9$}& &\multirow{12}{*}{$e\nu$} & \multirow{12}{*}{$86\pm12$} & \multirow{12}{*}{79}\\
&  & $\mu\nu$& {$0.0435\pm0.0034$}& & {$\phantom{0}17.9\pm1.4$}& & \multirow{12}{*}{$\mu\nu$}& \multirow{12}{*}{\phantom{00}$65.5\pm8.5$} & \multirow{12}{*}{40}\\
\multirow{2}{*}{100} & 
& $e\nu$ & {$0.0798\pm0.0050$}& & {$\phantom{0}31.0\pm1.9$}&  \\
&  & $\mu\nu$& {$0.0437\pm0.0034$}& & {$\phantom{0}17.0\pm1.3$}& \\
\multirow{2}{*}{200} &
 & $e\nu$ & {$0.0762\pm0.0049$}& & {$\phantom{0}25.1\pm1.6$}&  \\
&  & $\mu\nu$& {$0.0461\pm0.0034$}& & {$\phantom{0}15.2\pm1.1$}&  \\
\multirow{2}{*}{400} & 
& $e\nu$ & {$0.0857\pm0.0055$}& & {$\phantom{0}16.2\pm1.0$}&  \\
& & $\mu\nu$ & {$0.0532\pm0.0040$}& & {$\phantom{0}10.0\pm0.8$}&  \\
\multirow{2}{*}{1000} &
 & $e\nu$ & {$0.0987\pm0.0091$}& & {$\phantom{0}1.28\pm0.12$}&  \\
&  & $\mu\nu$& {$0.0636\pm0.0057$}& & {$\phantom{}0.824\pm0.074$}&  \\
\multirow{2}{*}{1300} & 
 & $e\nu$ & {$0.1010\phantom{}\pm0.0095$}& & {$\phantom{}0.240\pm0.023$}&  \\
&  & $\mu\nu$& {$0.0589\pm0.0057$}& & {$\phantom{}0.140\pm0.014$}&  \\
    \hline
    & & \multicolumn{8}{c}{D9 Operator}\\
    \hline
\multirow{2}{*}{1} &  \multirow{12}{*}{843}
 & $e\nu$ & {$0.0851\pm0.0053$}& & {$\phantom{0}55.5\pm3.5$}& & \multirow{12}{*}{$e\nu$}&\multirow{12}{*}{$86\pm12$} & \multirow{12}{*}{79}\\
&  & $\mu\nu$& {$0.0517\pm0.0035$}& & {$\phantom{0}33.8\pm2.3$}& &\multirow{12}{*}{$\mu\nu$}& \multirow{12}{*}{\phantom{00}$65.5\pm8.5$} & \multirow{12}{*}{40}\\
\multirow{2}{*}{100} &
 & $e\nu$ & {$0.0950\pm0.0056$}& & {$\phantom{0}55.8\pm3.3$}&  \\
&  & $\mu\nu$& {$0.0529\pm0.0038$}& & {$\phantom{0}31.1\pm2.3$}&  \\
\multirow{2}{*}{200} &
 & $e\nu$ & {$0.1040\phantom{}\pm0.0062$}& & {\phantom{0}$48.9\pm2.9$}&  \\
& & $\mu\nu$& {$0.0553\pm0.0039$}& & {$\phantom{0}26.0\pm1.8$}&  \\
\multirow{2}{*}{400} &
 & $e\nu$ & {$0.1030\phantom{}\pm0.0067$}& & {$\phantom{0}25.5\pm1.6$}&  \\
& & $\mu\nu$& {$0.0578\pm0.0042$}& & {$\phantom{0}14.3\pm1.0$}&  \\
\multirow{2}{*}{1000} & 
 & $e\nu$ & {$0.1070\pm0.0092$}& & {$\phantom{0}1.63\pm0.14$}&  \\
&  & $\mu\nu$& 0{$.0615\pm0.0055$}& & {$\phantom{}0.944\pm0.084$}&  \\
\multirow{2}{*}{1300} & 
 & $e\nu$ & {$0.1020\pm0.0100$}& & {$\phantom{}0.285\pm0.029$}&  \\
& & $\mu\nu$ & {$0.0573\pm0.0056$}& & {$\phantom{}0.160\pm0.016$}&  \\
\hline
\hline
\end{tabular}
\end{center}
\end{table*}

\FloatBarrier

The number of observed events is generally in good agreement with the expected number of background events for all
mass bins. None of the observations for any mass point in either channel or their combination show a significant
excess above background, so there is no evidence for the observation of either \wpl\ or \wsl. A deficit in the number of observed events
with respect to the expected number of background events is observed in the muon channel. This deficit has at most a 2.2$\sigma$ local significance. 

Tables~\ref{tab:limits_xbrfid_wp_and_ws1} and~\ref{tab:limits_xbrfid_wp_and_ws2} and figure~\ref{fig:final_limits} present the
95\% confidence level (CL) observed limits on \xbr\ for both \wpl\ and \wsl\
in the electron channel, the muon channel and their combination.
The tables also give the limits obtained without systematic uncertainties. 
Limits with various subsets of the systematic uncertainties are shown for \wpl\  as a representative case.
The uncertainties on the signal selection efficiency have very little effect on the final limits, and the
background-level and luminosity uncertainties are important only for the lowest masses.
Figure~\ref{fig:final_limits} also shows the expected limits and the theoretical \xbr\ for a \wp\ 
and for a \wstar. Limits are evaluated by fixing the \wstar\ coupling strengths to give the same partial decay
widths as the \wp. The off-shell production of \wp\ degrades the acceptance at high mass, 
worsening the limits. As discussed in chapter~\ref{sec:intro}, \wstar\ has different couplings with respect to \wp, 
enhancing the production at the pole. Since the off-shell production is reduced with respect to  \wp, 
the \wstar\ limits do not show the same behaviour at high mass.

\begin{table}[!htbp]
\caption{
  Observed upper limits on \xbr\  for \wp\  and \wstar\ with masses up to 2000 GeV. The first column is
    the \wps\ mass and the following columns refer to the 95\% CL limits for the \wp\ with
    headers indicating the nuisance parameters for which uncertainties
    are included: S for the event selection efficiency (\effsig), B
    for the background level (\nbg), and L for the integrated
    luminosity (\lint).  The column labelled SBL includes all
    uncertainties neglecting correlations.  Results are also presented when including the correlation of the
    signal and background cross-section uncertainties, as well as the correlation of the background cross-section uncertainties for the combined limits
    ($\mathrm{SB_c}$, $\mathrm{SB_cL}$).  The last two columns show 
    the limits for the $W^{*}$ without nuisance parameters and when including all nuisance parameters 
    with correlations.
\label{tab:limits_xbrfid_wp_and_ws1}
}
\begin{center}
\footnotesize
\begin{tabular}{cc|rrrrrr|rr}
\hline
\hline
   \mwps ~[\gev] &  Channel                & \multicolumn{8}{c}{95\% CL limit on  \xbr\ [fb]}   \\
                             &                                &  \multicolumn{6}{c}{\wp\ } &  \multicolumn{2}{c}{$W^{*}$} \\
                             &                                &none & S & SB  & SBL & $\mathrm{SB_c}$ & $\mathrm{SB_cL}$ & none & $\mathrm{SB_cL}$\\
    \hline
 \multirow{3}{*}{300}
 & $e\nu$     & 29.0  & 29.1  & 304  & 342  & 305  & 343  \\ 
 & $\mu\nu$ & 22.4  & 22.4  & 327  & 363  & 327  & 363  \\ 
 & both         & 14.2  & 14.2  & 219  & 269  & 290  & 331  \\ 
\hline 
\multirow{3}{*}{400}
 & $e\nu$ & 14.1  & 14.1  & 94.8  & 105  & 95.0  & 105  & 20.7  & 204  \\ 
 & $\mu\nu$ & 12.6  & 12.6  & 91.3  & 102  & 91.4  & 102  & 25.1  & 233  \\ 
 & both & 7.55  & 7.56  & 63.4  & 77.0  & 83.2  & 94.7  & 12.6  & 197  \\ 
\hline 
\multirow{3}{*}{500}
 & $e\nu$ & 9.14  & 9.18  & 38.7  & 42.2  & 38.8  & 42.4  & 17.3  & 87.5  \\ 
 & $\mu\nu$ & 6.42  & 6.44  & 30.6  & 34.0  & 30.7  & 34.1  & 10.5  & 77.9  \\ 
 & both & 4.26  & 4.26  & 22.3  & 27.0  & 29.8  & 33.9  & 7.54  & 77.7  \\ 
\hline 
\multirow{3}{*}{600}
 & $e\nu$ & 5.67  & 5.68  & 19.5  & 21.2  & 19.7  & 21.4  & 10.4  & 43.9  \\ 
 & $\mu\nu$ & 4.38  & 4.40  & 15.5  & 17.0  & 15.6  & 17.1  & 7.11  & 32.8  \\ 
 & both & 2.78  & 2.78  & 11.1  & 13.2  & 15.5  & 17.4  & 4.75  & 33.9  \\ 
\hline 
\multirow{3}{*}{750}
 & $e\nu$ & 2.95  & 2.95  & 8.25  & 8.71  & 8.35  & 8.81  & 4.23  & 14.9  \\ 
 & $\mu\nu$ & 3.33  & 3.34  & 7.89  & 8.35  & 7.97  & 8.43  & 5.23  & 14.7  \\ 
 & both & 1.73  & 1.73  & 5.06  & 5.63  & 7.01  & 7.52  & 2.51  & 12.8  \\ 
\hline 
\multirow{3}{*}{1000}
 & $e\nu$ & 1.84  & 1.85  & 3.25  & 3.34  & 3.29  & 3.38  & 2.69  & 6.01  \\ 
 & $\mu\nu$ & 1.86  & 1.87  & 2.87  & 2.95  & 2.92  & 3.00  & 3.02  & 5.88  \\ 
 & both & 1.03  & 1.04  & 1.86  & 1.96  & 2.48  & 2.58  & 1.57  & 4.94  \\ 
\hline 
\multirow{3}{*}{1250}
 & $e\nu$ & 1.63  & 1.64  & 2.06  & 2.09  & 2.09  & 2.12  & 2.29  & 3.65  \\ 
 & $\mu\nu$ & 1.62  & 1.62  & 2.01  & 2.04  & 2.04  & 2.07  & 1.78  & 2.60  \\ 
 & both & 0.990  & 0.991  & 1.30  & 1.34  & 1.54  & 1.57  & 1.16  & 2.53  \\ 
\hline 
\multirow{3}{*}{1500}
 & $e\nu$ & 1.27  & 1.28  & 1.40  & 1.41  & 1.42  & 1.43  & 1.99  & 2.39  \\ 
 & $\mu\nu$ & 1.21  & 1.22  & 1.35  & 1.36  & 1.37  & 1.38  & 1.71  & 2.06  \\ 
 & both & 0.775  & 0.777  & 0.879  & 0.890  & 0.967  & 0.979  & 1.14  & 1.63  \\ 
\hline 
\multirow{3}{*}{1750}
 & $e\nu$ & 0.964  & 0.967  & 0.993  & 0.997  & 1.01  & 1.01  & 1.48  & 1.64  \\ 
 & $\mu\nu$ & 0.813  & 0.818  & 0.818  & 0.821  & 0.827  & 0.831  & 1.37  & 1.54  \\ 
 & both & 0.521  & 0.522  & 0.533  & 0.537  & 0.563  & 0.567  & 0.889  & 1.10  \\ 
\hline 
\multirow{3}{*}{2000}
 & $e\nu$ & 0.721  & 0.724  & 0.735  & 0.738  & 0.743  & 0.746  & 1.34  & 1.40  \\ 
 & $\mu\nu$ & 0.747  & 0.751  & 0.751  & 0.754  & 0.760  & 0.762  & 1.18  & 1.26  \\ 
 & both & 0.415  & 0.416  & 0.422  & 0.424  & 0.439  & 0.441  & 0.831  & 0.922  \\ 
\hline 
\hline
\end{tabular}
\end{center}
\end{table}

\begin{table}[!htbp]
\caption{
  Observed upper limits on \xbr\  for \wp\  and \wstar\ with masses above 2000 GeV.
  The columns are the same as in table \ref{tab:limits_xbrfid_wp_and_ws1}.
  \label{tab:limits_xbrfid_wp_and_ws2}
}
\begin{center}
\footnotesize
\begin{tabular}{cc|rrrrrr|rr}
\hline
\hline
   \mwps ~[\gev] &  Channel                & \multicolumn{8}{c}{95\% CL limit on  \xbr\ [fb]}   \\
                             &                                &  \multicolumn{6}{c}{\wp\ } &  \multicolumn{2}{c}{$W^{*}$} \\
                             &                                &none & S & SB  & SBL & $\mathrm{SB_c}$ & $\mathrm{SB_cL}$ & none & $\mathrm{SB_cL}$\\
    \hline
 \multirow{3}{*}{2250}
 & $e\nu$ & 0.453  & 0.455  & 0.455  & 0.456  & 0.458  & 0.459  & 0.830  & 0.859  \\ 
 & $\mu\nu$ & 0.853  & 0.859  & 0.859  & 0.862  & 0.866  & 0.869  & 0.726  & 0.734  \\ 
 & both & 0.296  & 0.297  & 0.297  & 0.298  & 0.301  & 0.303  & 0.457  & 0.488  \\ 
\hline 
\multirow{3}{*}{2500}
 & $e\nu$ & 0.564  & 0.569  & 0.569  & 0.570  & 0.572  & 0.573  & 0.438  & 0.441  \\ 
 & $\mu\nu$ & 1.06  & 1.07  & 1.07  & 1.08  & 1.08  & 1.08  & 0.828  & 0.837  \\ 
 & both & 0.368  & 0.370  & 0.370  & 0.371  & 0.376  & 0.377  & 0.287  & 0.289  \\ 
\hline 
\multirow{3}{*}{2750}
 & $e\nu$ & 0.629  & 0.643  & 0.643  & 0.644  & 0.648  & 0.649  & 0.459  & 0.462  \\ 
 & $\mu\nu$ & 1.16  & 1.19  & 1.19  & 1.20  & 1.21  & 1.21  & 0.917  & 0.928  \\ 
 & both & 0.409  & 0.413  & 0.413  & 0.414  & 0.425  & 0.426  & 0.306  & 0.308  \\ 
\hline 
\multirow{3}{*}{3000}
 & $e\nu$ & 0.809  & 0.852  & 0.852  & 0.853  & 0.863  & 0.865  & 0.387  & 0.389  \\ 
 & $\mu\nu$ & 1.47  & 1.55  & 1.55  & 1.56  & 1.58  & 1.58  & 0.798  & 0.807  \\ 
 & both & 0.523  & 0.534  & 0.534  & 0.536  & 0.566  & 0.567  & 0.261  & 0.263  \\ 
\hline 
\multirow{3}{*}{3250}
 & $e\nu$ & 1.20  & 1.37  & 1.37  & 1.37  & 1.40  & 1.40  & 0.338  & 0.340  \\ 
 & $\mu\nu$ & 2.14  & 2.45  & 2.45  & 2.45  & 2.52  & 2.52  & 0.678  & 0.687  \\ 
 & both & 0.768  & 0.815  & 0.815  & 0.816  & 0.919  & 0.920  & 0.226  & 0.228  \\ 
\hline 
\multirow{3}{*}{3500}
 & $e\nu$ & 1.92  & 2.56  & 2.56  & 2.56  & 2.64  & 2.64  & 0.312  & 0.315  \\ 
 & $\mu\nu$ & 3.37  & 4.38  & 4.38  & 4.39  & 4.56  & 4.57  & 0.645  & 0.655  \\ 
 & both & 1.22  & 1.38  & 1.38  & 1.38  & 1.72  & 1.73  & 0.210  & 0.213  \\ 
\hline 
\multirow{3}{*}{3750}
 & $e\nu$ & 3.12  & 4.90  & 4.90  & 4.90  & 5.07  & 5.08  & 0.297  & 0.307  \\ 
 & $\mu\nu$ & 5.32  & 7.85  & 7.85  & 7.86  & 8.22  & 8.24  & 0.605  & 0.630  \\ 
 & both & 1.97  & 2.37  & 2.37  & 2.38  & 3.26  & 3.27  & 0.199  & 0.208  \\ 
\hline 
\multirow{3}{*}{4000}
 & $e\nu$ & 4.76  & 8.07  & 8.07  & 8.09  & 8.38  & 8.40  & 0.304  & 0.372  \\ 
 & $\mu\nu$ & 7.75  & 12.0  & 12.0  & 12.0  & 12.6  & 12.6  & 0.613  & 0.749  \\ 
 & both & 2.95  & 3.66  & 3.66  & 3.66  & 5.24  & 5.24  & 0.203  & 0.255  \\ 
  \hline
\hline
\end{tabular}
\end{center}
\end{table}

\begin{figure*}[!htbp]
  \centering
  \includegraphics[width=0.49\textwidth]{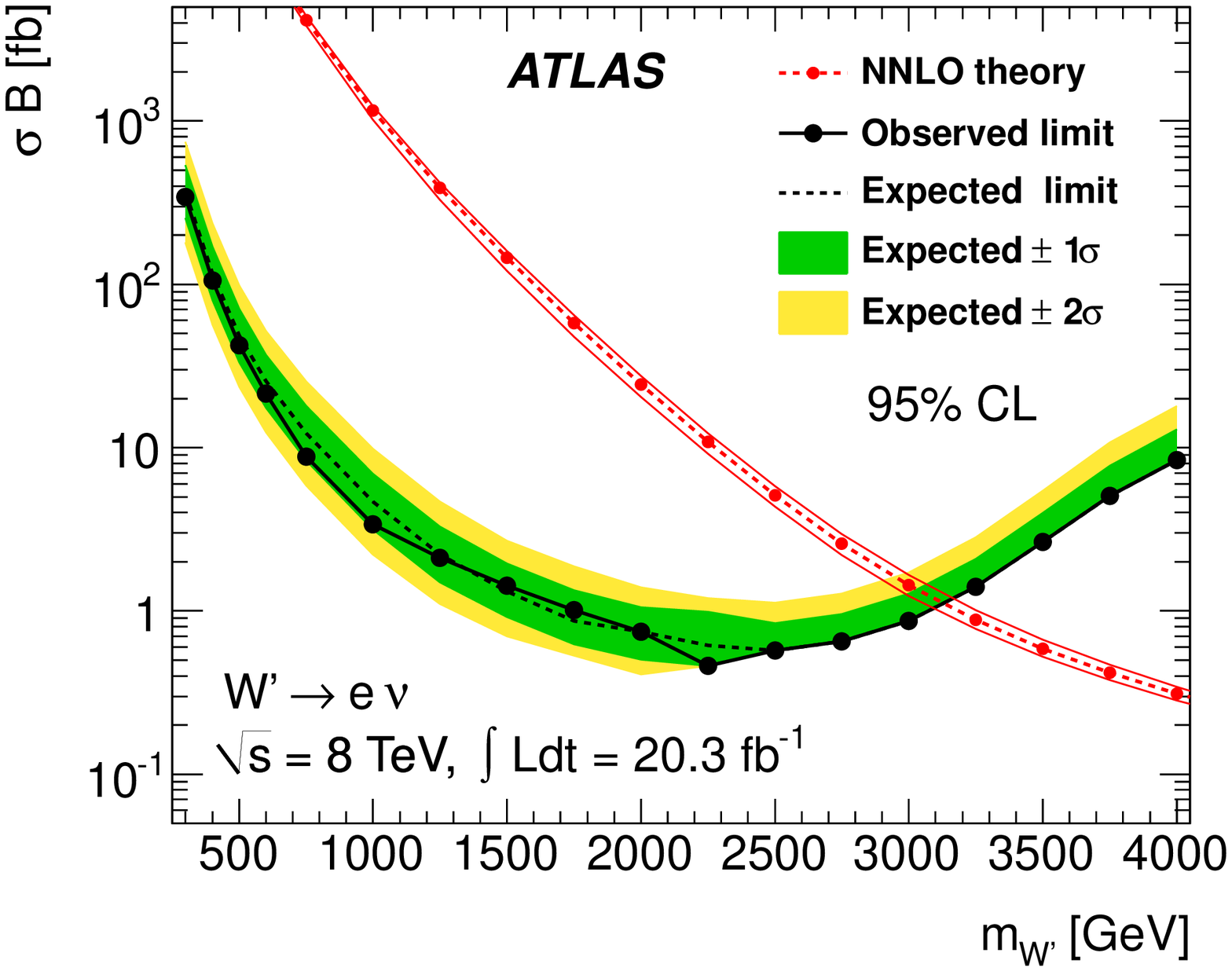}
  \includegraphics[width=0.49\textwidth]{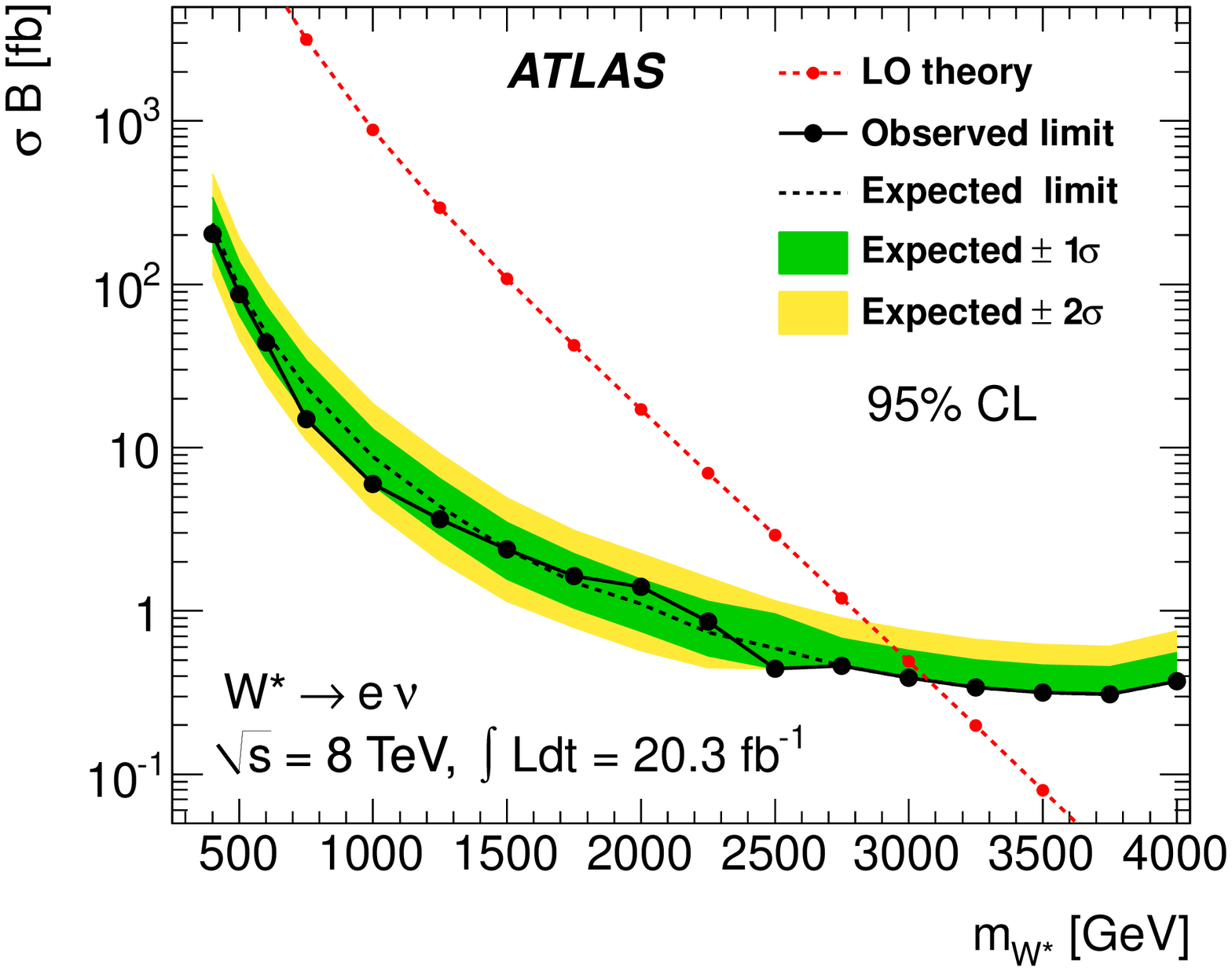}
  \includegraphics[width=0.49\textwidth]{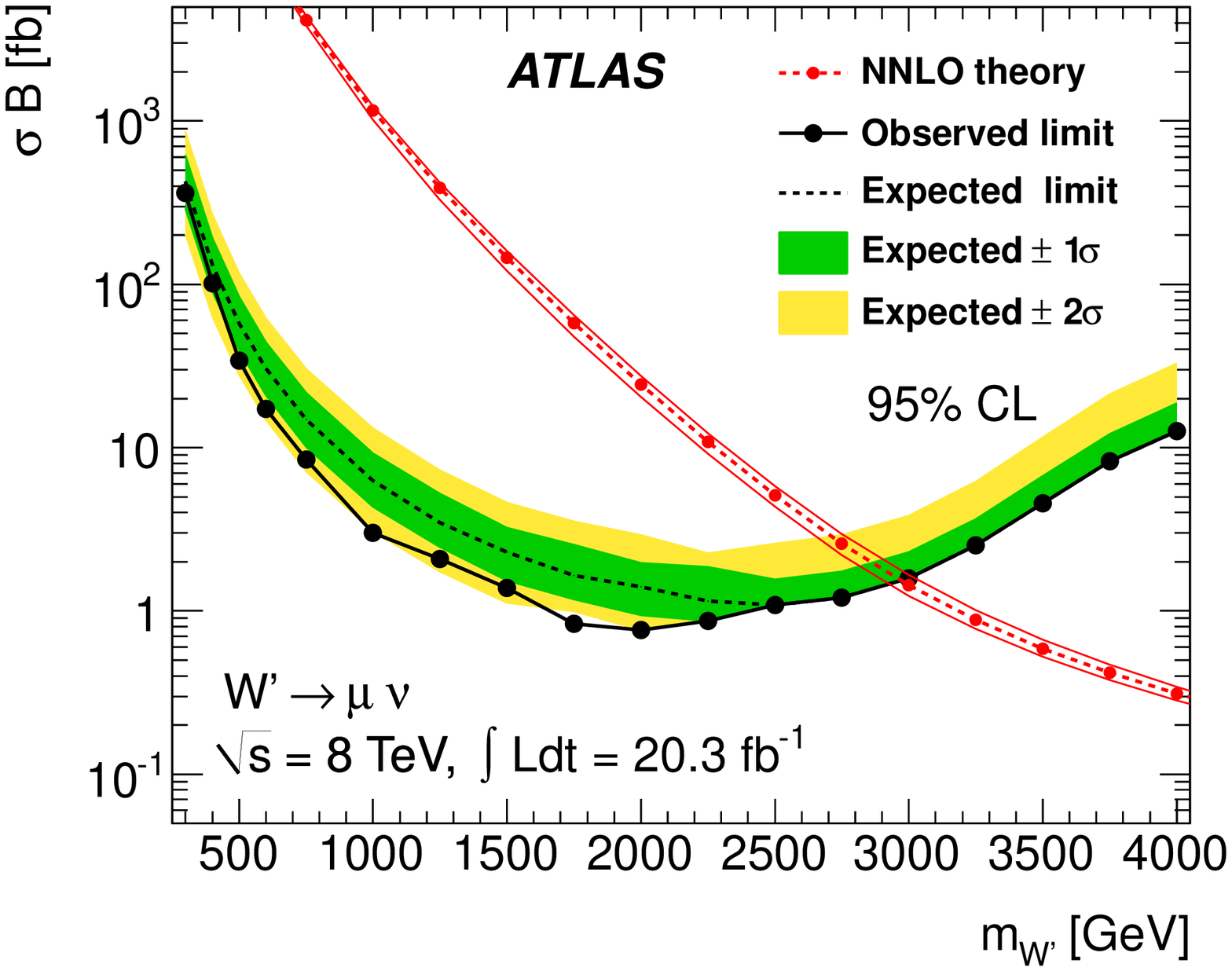}
  \includegraphics[width=0.49\textwidth]{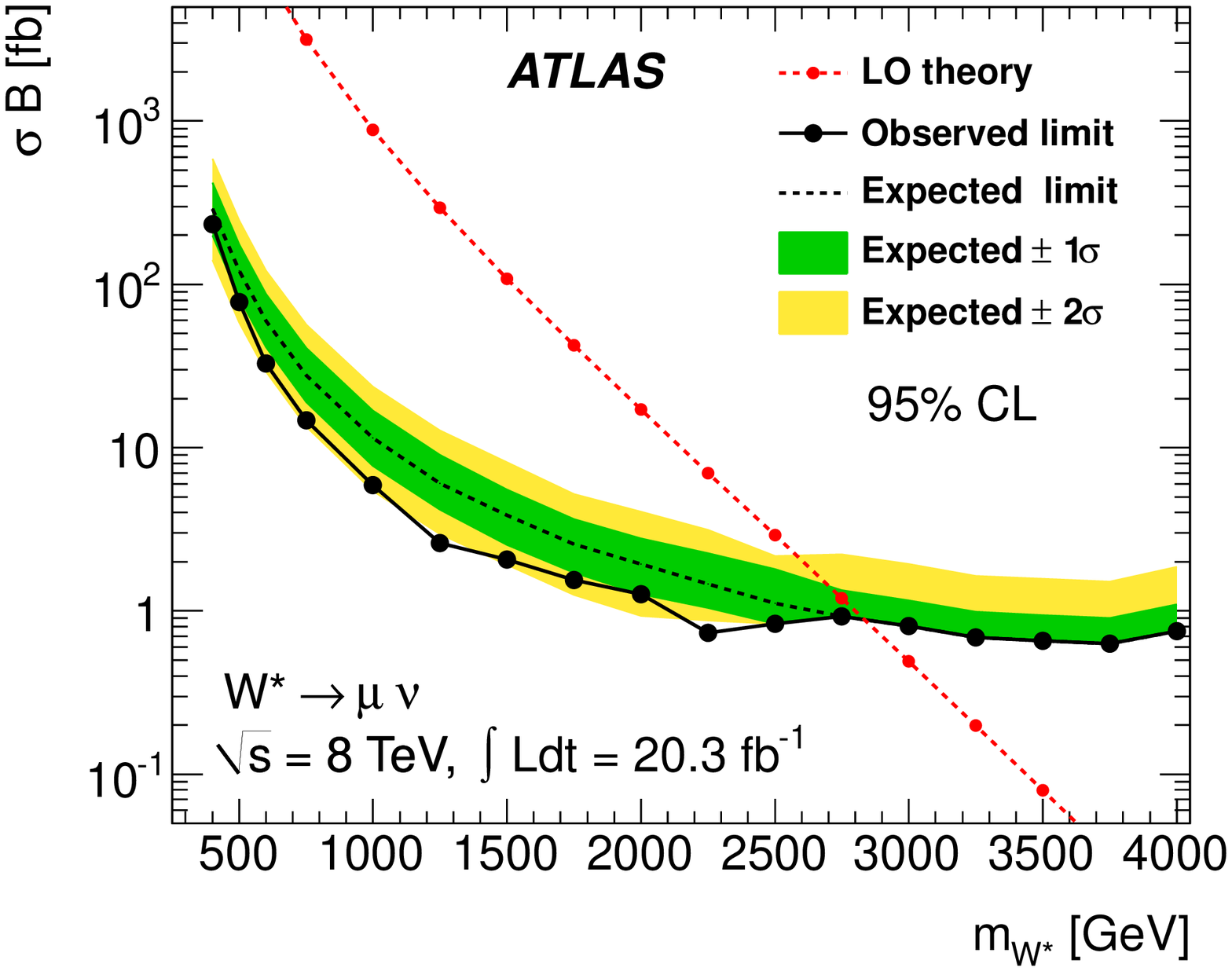}
  \includegraphics[width=0.49\textwidth]{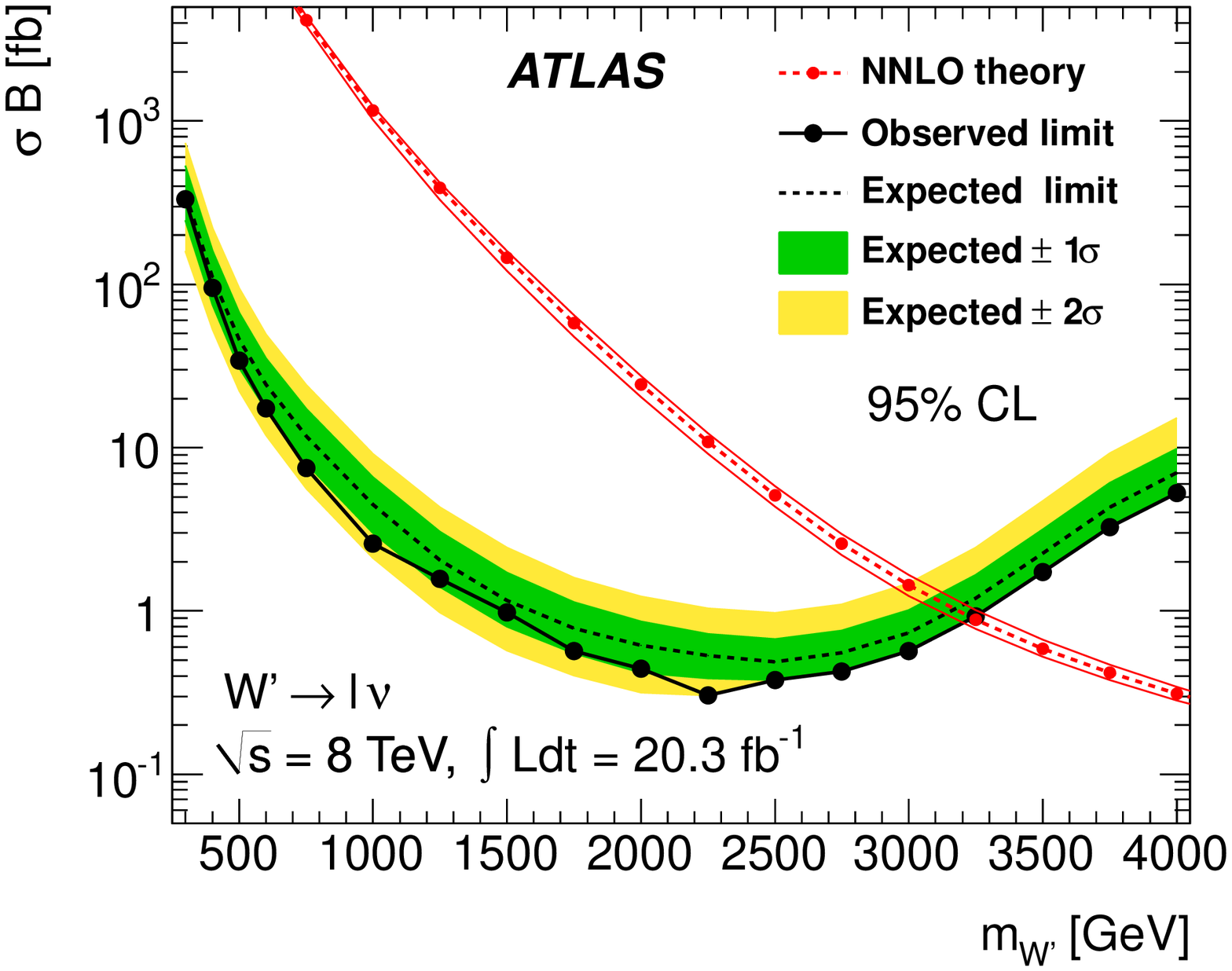}
  \includegraphics[width=0.49\textwidth]{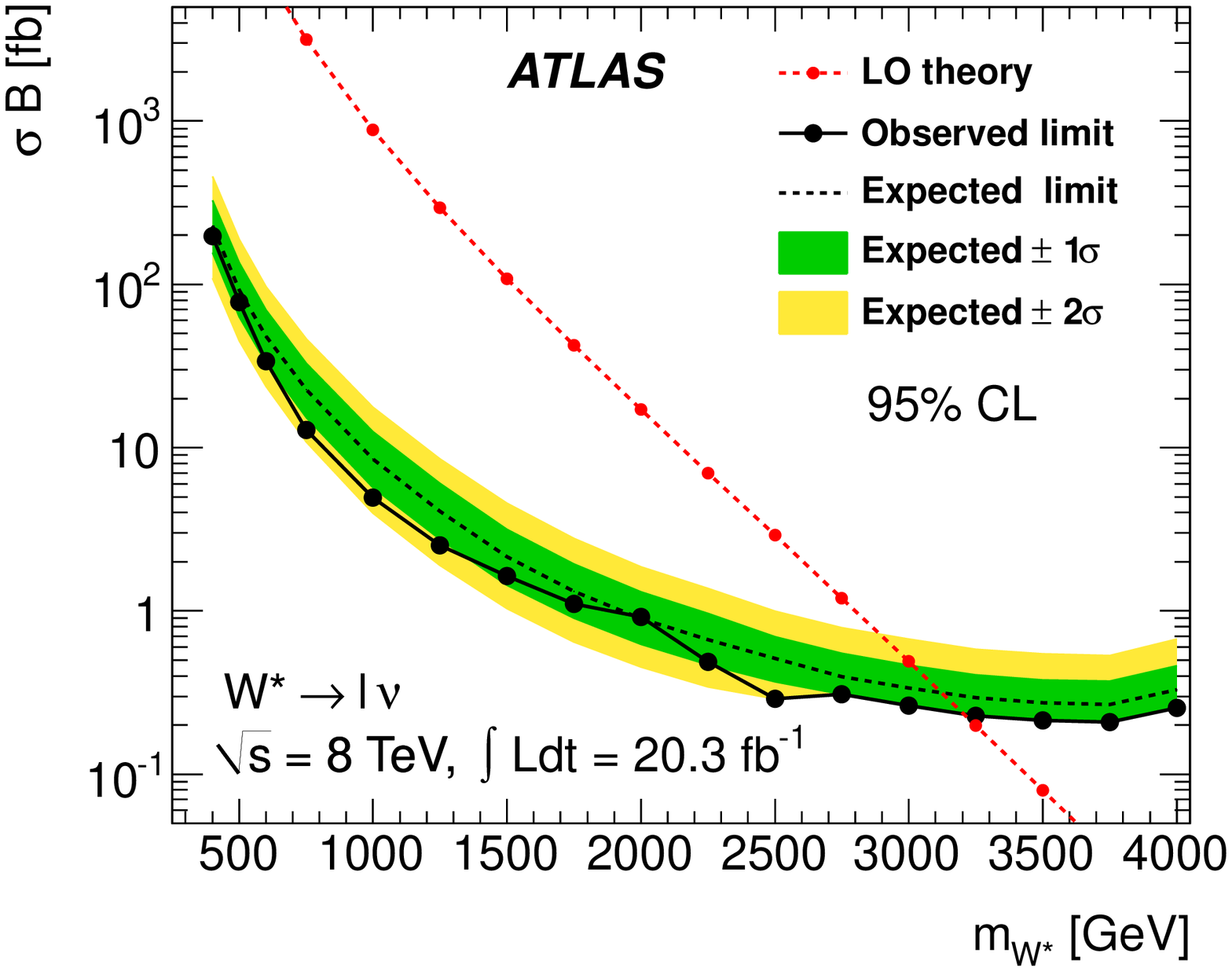}
  \caption{Observed and expected  limits on \xbr\ for \wp\ (left) and $W^{*}$ (right) at 95\% CL in the electron
  channel (top), muon channel (centre) and the combination (bottom) assuming the same branching fraction for both channels. 
  The predicted values for \xbr\  and their uncertainties (except for $W^{*}$) are also shown.  
  The calculation of uncertainties on the \wp\  cross-sections is explained in section \ref{sec:mcSim}.
  \label{fig:final_limits}
  }
\end{figure*}

\begin{table}[!htbp]
\caption{Lower limits on the \wp\ and $W^{*}$ masses. The
    first column is the decay channel ($e\nu$, $\mu\nu$ or both
    combined) and the following give the expected (Exp.) and observed
    (Obs.) mass limits.
  \label{tab:limits_mass_wp_and_ws}
}
 \vspace{3 mm}
  \centering
  \begin{tabular}{c|rr|rr}
    \hline
    \hline
    &  \multicolumn{2}{c}{\mwp\ [\tev]}  & \multicolumn{2}{c}{\mws\ [\tev]} \\
   Decay     &  Exp. & Obs.  & Exp. & Obs. \\
  \hline
  $e\nu$        & 3.13 & 3.13 & 3.08 & 3.08 \\
  $\mu\nu$      & 2.97 & 2.97 & 2.83 & 2.83 \\
  Both           & 3.17 & 3.24 & 3.12 & 3.21 \\
       \hline
    \hline
  \end{tabular}
 \end{table}
 
In figure~\ref{fig:final_limits} the intersection between the central theoretical prediction and 
the observed limits provides the 95\% CL lower limits on the mass.
The expected and observed \wp\  and \wstar\ mass limits for the electron and muon decay channels 
as well as their combination are listed in table~\ref{tab:limits_mass_wp_and_ws}.  
The  difference between the expected and observed combined mass limits originate from the
slight data deficit in each decay channel that are individually not significant.
The band around the theoretical prediction in figure~\ref{fig:final_limits} indicates the total theory uncertainty as described earlier in the text.
The mass limit for the \wp\ decreases by 50 \gev\ if the intersection between the lower theoretical prediction and the observed limit is used. 
The uncertainties on \effsig, \nbg\ and \lint\ affect the derived mass limits by a similar amount.
Limits are also evaluated following the $\mathrm{CL_{s}}$ prescription~\cite{aread} using the profile likelihood ratio as the test statistic including
all uncertainties. The cross-section limits are found to agree within 10\% across the entire mass range, with only marginal impact on the mass limit.
The mass limits presented here are a significant improvement over those reported in previous ATLAS and CMS searches 
\cite{atlas:wprime_2010_pub,atlas:wprime_2011-2_pub,atlas:wprime_2012_pub,cms:wprime2013-1}.

The results of the search for pair production of DM particles in association with a leptonically decaying $W$ boson 
are shown in figures~\ref{fig:mstar_vs_mchi} and~\ref{fig:chi_nucleon_xsec}.
The former shows the observed limits on $M_{*}$, the mass scale of the unknown mediating interaction for the DM particle pair production,
whereas the latter shows the observed limits on the DM--nucleon scattering cross-section. Both are shown as a function of the DM particle mass, $m_{\chi}$, 
and presented at 90\% CL. Results of the previous ATLAS searches for hadronically decaying $W/Z$~\cite{atlasWIMPhadronicWZ}, leptonically decaying 
$Z$~\cite{atlasWIMPleptonicZ}, and $j+\chi\chi$~\cite{atlasWIMPjet} are also shown. The observed limits on $M_{*}$  as a function of  $m_{\chi}$
are by a factor $\sim$1.5 stronger in the search for DM production in  association with hadronically decaying $W$ with respect the ones
presented in this paper.

\begin{figure*}[!htbp]
  \centering
   \includegraphics[width=0.8\textwidth]{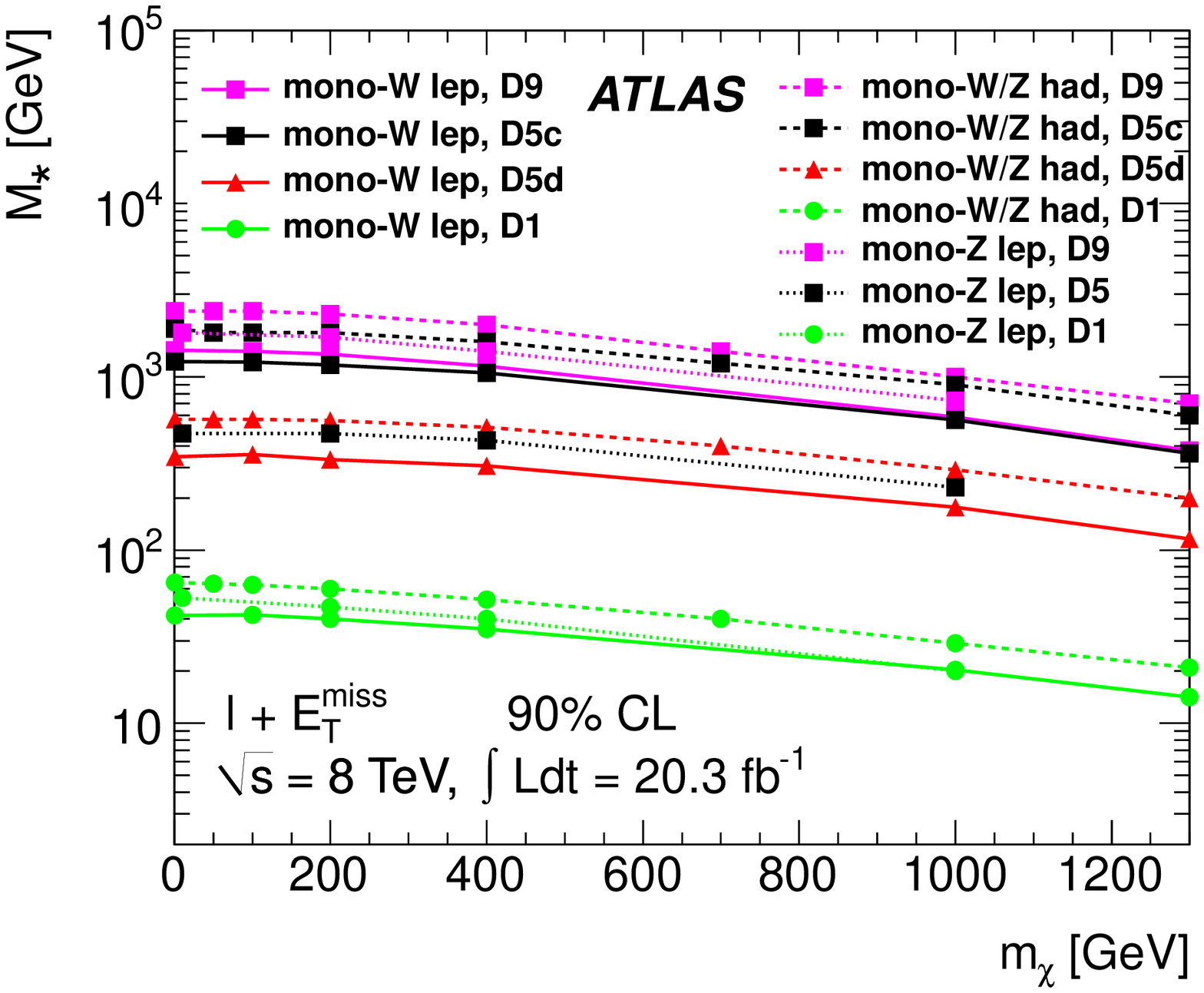}
  \caption{Observed limits on $M_{*}$ as a function of the DM particle mass ($m_{\chi}$) at 90\% CL 
  for the combination of the electron and muon channel,
  for various operators as described in the text. 
  For each operator, the values below the corresponding line are excluded. 
  No signal samples are generated for masses below 1 \GeV\ but the limits are expected to be stable down to arbitrarily small values.
  Results of the previous ATLAS searches for hadronically decaying $W/Z$~\cite{atlasWIMPhadronicWZ} and 
  leptonically decaying $Z$~\cite{atlasWIMPleptonicZ} are also shown.
  \label{fig:mstar_vs_mchi}
  }
\end{figure*}

\begin{figure*}[!htbp]
	\centering
	\includegraphics[width=0.9\textwidth]{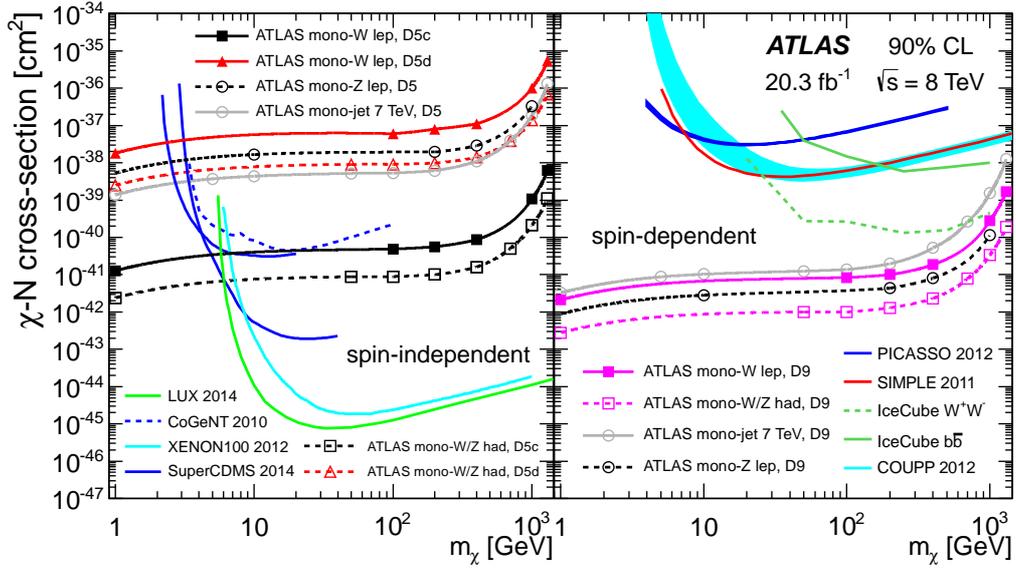}
	\caption{Observed limits on the DM--nucleon scattering cross-section as a function of $m_{\chi}$ 
	at 90\% CL for spin-independent (left) and spin-dependent (right) operators in the EFT. 
	Results are compared with the previous ATLAS searches for hadronically decaying $W/Z$~\cite{atlasWIMPhadronicWZ},
	leptonically decaying $Z$~\cite{atlasWIMPleptonicZ}, and $j+\chi\chi$~\cite{atlasWIMPjet}, and with direct detection searches by
	CoGeNT~\cite{2011PhRvL.106m1301A}, XENON100~\cite{2012PhRvL.109r1301A}, 
	CDMS~\cite{2014PhRvL.112d1302A,2014arXiv1402.7137A}, LUX~\cite{2013arXiv1310.8214L},
	COUPP~\cite{2012PhRvD..86e2001B}, SIMPLE~\cite{2012PhRvL.108t1302F},
	PICASSO~\cite{2012PhLB..711..153A} and IceCube~\cite{2012arXiv1212.4097I}.
	The comparison between direct detection and ATLAS results is only possible within the limits of the 
	validity of the EFT~\cite{wimp14TeV}.
	\label{fig:chi_nucleon_xsec}}
\end{figure*}

\FloatBarrier

\section{Conclusions}
\label{sec:conclusions}

A search is presented for new high-mass states decaying to a lepton (electron or muon) 
plus missing transverse momentum using 20.3~\ifb\  of proton--proton collision data at  
$\sqrt{s}=8\tev$ recorded with the ATLAS experiment at the Large Hadron Collider.
No significant excess beyond SM expectations is observed. Limits on \xbr\ are presented. 
A \wp\ with SSM couplings is excluded for masses below 3.24~\tev\ at 95\% CL.  
The exclusion for \wstar\ with equivalent couplings is 3.21~\tev.  
For the pair production of weakly interacting  DM particles in events with a leptonically decaying $W$, limits are set on 
the mass scale, \mstar, of the unknown mediating interaction as well as on the DM--nucleon scattering cross-section.

\FloatBarrier

\section*{Acknowledgements}
\label{sec:acknowled}

We thank CERN for the very successful operation of the LHC, as well as the
support staff from our institutions without whom ATLAS could not be
operated efficiently.

We acknowledge the support of ANPCyT, Argentina; YerPhI, Armenia; ARC,
Australia; BMWF and FWF, Austria; ANAS, Azerbaijan; SSTC, Belarus; CNPq and FAPESP,
Brazil; NSERC, NRC and CFI, Canada; CERN; CONICYT, Chile; CAS, MOST and NSFC,
China; COLCIENCIAS, Colombia; MSMT CR, MPO CR and VSC CR, Czech Republic;
DNRF, DNSRC and Lundbeck Foundation, Denmark; EPLANET, ERC and NSRF, European Union;
IN2P3-CNRS, CEA-DSM/IRFU, France; GNSF, Georgia; BMBF, DFG, HGF, MPG and AvH
Foundation, Germany; GSRT and NSRF, Greece; ISF, MINERVA, GIF, I-CORE and Benoziyo Center,
Israel; INFN, Italy; MEXT and JSPS, Japan; CNRST, Morocco; FOM and NWO,
Netherlands; BRF and RCN, Norway; MNiSW and NCN, Poland; GRICES and FCT, Portugal; MNE/IFA, Romania; 
MES of Russia and ROSATOM, Russian Federation; JINR; MSTD,
Serbia; MSSR, Slovakia; ARRS and MIZ\v{S}, Slovenia; DST/NRF, South Africa;
MINECO, Spain; SRC and Wallenberg Foundation, Sweden; SER, SNSF and Cantons of
Bern and Geneva, Switzerland; NSC, Taiwan; TAEK, Turkey; STFC, the Royal
Society and Leverhulme Trust, United Kingdom; DOE and NSF, United States of
America.

The crucial computing support from all WLCG partners is acknowledged
gratefully, in particular from CERN and the ATLAS Tier-1 facilities at
TRIUMF (Canada), NDGF (Denmark, Norway, Sweden), CC-IN2P3 (France),
KIT/GridKA (Germany), INFN-CNAF (Italy), NL-T1 (Netherlands), PIC (Spain),
ASGC (Taiwan), RAL (UK) and BNL (USA) and in the Tier-2 facilities
worldwide.

\bibliographystyle{JHEP}
\bibliography{references}

\clearpage
\onecolumn
\input{atlas_authlist}

\end{document}

%% file: atlas_authlist.tex
\begin{flushleft}
{\Large The ATLAS Collaboration}

\bigskip

G.~Aad$^{\rm 84}$,
B.~Abbott$^{\rm 112}$,
J.~Abdallah$^{\rm 152}$,
S.~Abdel~Khalek$^{\rm 116}$,
O.~Abdinov$^{\rm 11}$,
R.~Aben$^{\rm 106}$,
B.~Abi$^{\rm 113}$,
M.~Abolins$^{\rm 89}$,
O.S.~AbouZeid$^{\rm 159}$,
H.~Abramowicz$^{\rm 154}$
% ...
\bigskip
\\
% ...
$^{11}$ Institute of Physics, Azerbaijan Academy of Sciences, Baku, Azerbaijan\\
% ...
$^{84}$ CPPM, Aix-Marseille Universit{\'e} and CNRS/IN2P3, Marseille, France\\
% ...
$^{89}$ Department of Physics and Astronomy, Michigan State University, East Lansing MI, United States of America\\
% ...
$^{106}$ Nikhef National Institute for Subatomic Physics and University of Amsterdam, Amsterdam, Netherlands\\
% ...
$^{112}$ Homer L. Dodge Department of Physics and Astronomy, University of Oklahoma, Norman OK, United States of America\\
$^{113}$ Department of Physics, Oklahoma State University, Stillwater OK, United States of America\\
% ...
$^{116}$ LAL, Universit{\'e} Paris-Sud and CNRS/IN2P3, Orsay, France\\
% ...
$^{152}$ Institute of Physics, Academia Sinica, Taipei, Taiwan\\
% ...
$^{154}$ Raymond and Beverly Sackler School of Physics and Astronomy, Tel Aviv University, Tel Aviv, Israel\\
% ...
$^{159}$ Department of Physics, University of Toronto, Toronto ON, Canada\\
% ...
\end{flushleft}
